# Coherent X-ray Diffraction Imaging of Nanostructures

Ivan A. Vartanyants[1,2] and Oleksandr M. Yefanov[1]

[1]*Deutsches Elektronen Synchrotron DESY, Hamburg, Germany*
[2]*National Research Nuclear University, "MEPhI", Moscow, Russia*

## ABSTRACT

We present here an overview of Coherent X-ray Diffraction Imaging (CXDI) with its application to nanostructures. This imaging approach has become especially important recently due to advent of X-ray Free-Electron Lasers (XFEL) and its applications to the fast developing technique of serial X-ray crystallography. We start with the basic description of coherent scattering on the finite size crystals. The difference between conventional crystallography applied to large samples and coherent scattering on the finite size samples is outlined. The formalism of coherent scattering from a finite size crystal with a strain field is considered. Partially coherent illumination of a crystalline sample is developed. Recent experimental examples demonstrating applications of CXDI to the study of crystalline structures on the nanoscale, including experiments at FELs, are also presented.

## 1. INTRODUCTION

Coherent X-ray Diffractive Imaging (CXDI) is a relatively novel imaging method that can produce an image of a sample without using optics between the sample and detector (see Fig. 1). This differs from conventional microscopy schemes which use objective lenses to produce an image of an object. Taking into account the difficulties of producing lenses at hard X-ray energies that are both highly resolving and efficient, we see clearly the advantages of so-called 'lensless' microscopy techniques. After its first demonstration [1-4] CXDI was successfully applied at $3^{rd}$ generation synchrotron sources for imaging micron and nanometer size samples (see for recent reviews [5-10]).

The conventional CXDI experiment is performed with an isolated sample illuminated by a coherent, plane wave (Fig. 1). The incident wave may be described by a complex field (with a real and an imaginary part) of uniform magnitude and phase. The radiation interacts with the sample, which affects both the amplitude and phase of this field. The scattered radiation from the sample propagates to a two-dimensional detector in the far-field, and the diffracted intensities are measured. The detector can be positioned either in the forward direction, or in the case of a crystalline sample at Bragg angle positions (Fig. 1). It will be shown in the following sections that in the limit of kinematical scattering, which is a good approximation for scattering on nanostructures, the amplitude of the scattered field can be expressed as the Fourier transform (FT) of the electron density of a sample. However, the measurement of diffracted intensities exclusively is insufficient to unambiguously determine the electron density of a sample, as the phase information is lost during the measurement process (the measured quantity is the intensity and not the complex amplitude). Fortunately, with some additional knowledge of constraints on the sample in object space, the structure of the sample can be reconstructed using phase retrieval algorithms based on an iterative approach [11-13] (see also for a review of iterative methods [14]).

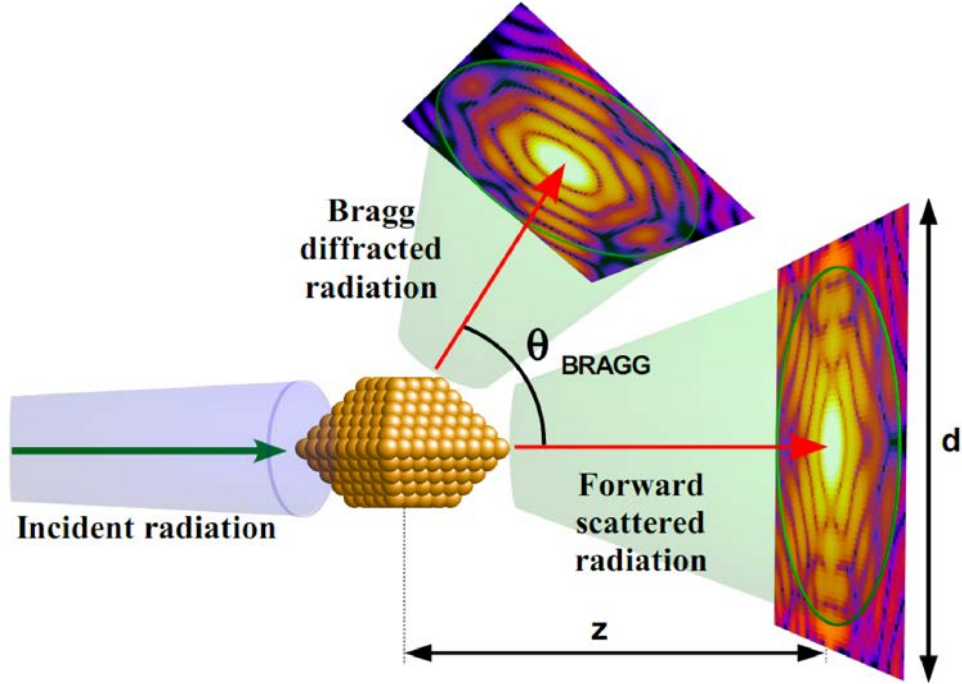

**Fig. 1** Schematic of a Coherent X-ray Diffractive Imaging (CXDI) experiment. Incident radiation illuminates a crystalline sample of a finite size (from the left). The forward scattered and diffracted radiation are measured in the far-field at a distance $z$ by an area detector of size $d$. The detector can be positioned either downstream from the sample position or at some Bragg angle position.

More formally, in CXDI experiments the modulus squared of the scattered amplitude, $A(\mathbf{q})$, where $\mathbf{q}$ is the momentum transfer, is measured. Note that the amplitude $A(\mathbf{q})$ is a complex function, while we can only measure its modulus squared. The image reconstruction process begins with assigning random phases to the known magnitudes $|A^{\exp}(\mathbf{q})| = \sqrt{I^{\exp}(\mathbf{q})}$ in reciprocal space. This complex wave in the far-field $A'(\mathbf{q}) = |A^{\exp}(\mathbf{q})| e^{i\phi(\mathbf{q})}$ is then inverse Fourier transformed to real space giving the first guess of the object in real space, $s'(\mathbf{r})$. This first guess will be unlike the correct solution and constraints in the object space, most importantly the finite extent of the object, have to be taken into account for better solutions. This typically involves setting the values of $s'(\mathbf{r})$ outside some bound to zero, known as the Error Reduction (ER) method, or forcing them towards zero, most commonly the Hybrid Input-Output (HIO) method [12]. After the constraints have been applied, the updated function $s(\mathbf{r})$ is then Fourier transformed to the far-field. The magnitude $|A(\mathbf{q})|$ of this new far-field guess is replaced by the measured intensities $|A^{\exp}(\mathbf{q})| = \sqrt{I^{\exp}(\mathbf{q})}$ while the phases $\phi(\mathbf{q})$ are kept. This process is then repeated for typically thousands of iterations until it converges (see Fig. 2). The resulting function $s(\mathbf{r})$ in kinematical approximation is the electron density of the sample.

A necessary condition for the successful reconstruction of the electron density of a sample from a diffraction pattern is the appropriate sampling of the pattern [15]. A useful experimental rule of thumb is that at least two measurement points per fringe in the diffraction pattern are required for adequate sampling. This means that the autocorrelation function of the data is correctly sampled according to Shannon's sampling theorem [16], which is twice what is required to sample the fully complex wave field. This essential sampling consideration leads this method to be sometimes referred to as the 'oversampling' method. We discuss these sampling conditions with application to reconstruction of crystalline structures in more detail in the following sections.

There are many variations of the standard iterative methods, one of which is known as the Guided Hybrid Input-Output (GHIO) method [17]. At this method $N$ different HIO reconstructions (called families) are performed in parallel, each with different random starting phases and typically continuing for a few thousand iterations. Subsequently, convergence criteria are applied to the results and the best reconstruction is used to seed another set of reconstructions through the expression $s^{g+1,n}(\mathbf{r}) = (s^{g,template}(\mathbf{r}) \times s^{g,n}(\mathbf{r}))^{1/2}$, where $n$ denotes a given family in the reconstruction and $g$ is termed the generation. The resulting $N$ terms $s^{g+1,n}(\mathbf{r})$ are used as inputs for the next 'generation' of HIO reconstructions. This process is repeated a number of times until convergence.

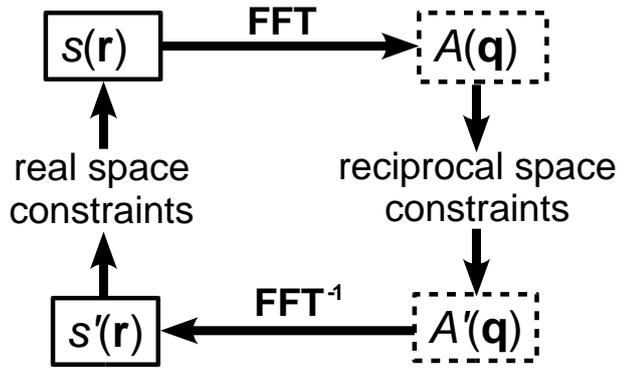

**Fig. 2** Block diagram of the iterative reconstruction method.

One of the strong limitations of conventional CXDI is the restriction of imaging only very small objects. A modification of the CXDI method known as 'ptychography' [18] removes this limitation. The ptychography is similar to scanning microscopy, where the sample is scanned through the X-ray beam and a far-field diffraction pattern is collected from each point of the sample scanned. Importantly, the illuminated regions should overlap in the sample plane. The relationship between each scan point is used to reconstruct the exit surface wave. This algorithm requires an overlap of the single illumination areas to yield a more stable reconstruction, and, most importantly, to avoid the finite object constraint of classical CXDI. A successful ptychographic phase and amplitude reconstruction of both the probe and an artificial gold zone plate to a resolution of tens of nanometers has been published [19] (see also [20]). The ability to image a field of view larger than the probe makes this an attractive technique for future application at hard condensed matter [21,22] and biological [23,24] systems.

As stated above, CXDI requires no optics between the sample and the detector which this is a great experimental simplification. However, there are also intrinsic limitations of this method. Typically, a CXDI iterative procedure requires some thousands of iterations for an image to be reconstructed, meaning that the imaging is certainly not performed instantaneously. Moreover, CXDI is a photon hungry method requiring increasingly higher flux to achieve higher resolutions. The reason is, that the scattered intensity is a function of momentum transfer and scales as $I(q) \propto q^{-k}$, where $k$ is between 3 and 4 depending upon the sample [25-27]. This means that for a generic object, a factor of three or four orders of magnitude in flux is required to improve the resolution by a single order, hence very bright photon beams are needed. Furthermore, a beam stop or a hole in the detector is required to prevent damage through the intense direct beam or the strong beam scattered at Bragg angles. This means a lack of data. Consequently, if this missing data region is too large, iterative reconstructions may fail [28]. Finally, the beam needs to be sufficiently coherent over the sample area, both transversely and longitudinally, for a successful reconstruction [29-31].

The paper is organized as follows. In the next section the basics of coherent scattering on a finite size crystal will be discussed. The difference between conventional crystallography applied to large samples and coherent scattering on a finite size samples will be outlined. Then the formalism of coherent scattering from a finite size crystal with a strain field will be developed. In the following section coherent scattering conditions will be relaxed and partially coherent illumination of a crystalline sample will be considered. In the last section some recent experimental examples demonstrating applications of CXDI to the study of crystalline structures on the nanoscale will be given.

## 2. COHERENT AND PARTIALLY COHERENT SCATTERING ON CRYSTALS

The scattering from an isotropic sample and a periodic crystal, when the intensity peaks at the Bragg positions, is quite different. R. Millane [32] was first to discuss in detail similarities and differences of the phase retrieval problem in crystallography and optics. Here we will give a short overview of this problem extending it to the case of finite size and strained crystals (see also [29]) where the link with the phase retrieval applied to non-crystallographic samples will be most evident. We will also discuss how this problem is connected with Shannon's sampling theorem and the possibilities of using the results of this theorem for phase retrieval in crystallography.

### *2.1 Coherent scattering from a finite size crystal*

It is well known (see for example [33,34]) that the scattering amplitude $A(\mathbf{q})$ of coherent monochromatic radiation from an infinite crystal in kinematical approximation[1] is equal to

$$A(\mathbf{q}) = \int \rho(\mathbf{r}) e^{-i\mathbf{q}\cdot\mathbf{r}} d^3r , \qquad (1)$$

where $\rho(\mathbf{r})$ is the electron density at the point $\mathbf{r}$, $\mathbf{q} = \mathbf{k}_f - \mathbf{k}_i$ is the momentum transfer and $\mathbf{k}_i$ and $\mathbf{k}_f$ are the incident and scattered wave vectors ($|\mathbf{k}_i| = |\mathbf{k}_f| = 2\pi/\lambda$, $\lambda$ is the wavelength). The electron density of a finite size crystal can be written as

$$\rho(\mathbf{r}) = \rho_{uc}(\mathbf{r}) \otimes [\rho_{\infty}(\mathbf{r}) \cdot s(\mathbf{r})] \quad , \tag{2}$$

where the sign $\otimes$ denotes the convolution. Here, $\rho_{uc}(\mathbf{r})$ is the electron density of a unit cell

$$\rho_{uc}(\mathbf{r}) = \sum_j \rho_j(\mathbf{r} - \mathbf{r}_j) \quad ,$$

where $\mathbf{r}_j$ is a coordinate and $\rho_j(\mathbf{r})$ is the electron density of individual atoms in a unit cell. To define an infinite ideal lattice we introduce the periodic function

$$\rho_{\infty}(\mathbf{r}) = \sum_{n=1}^{\infty} \delta(\mathbf{r} - \mathbf{R}_n) \quad ,$$

where $\mathbf{R}_n = n_1 \mathbf{a}_1 + n_2 \mathbf{a}_2 + n_3 \mathbf{a}_3$ is the position of the unit cell and $\mathbf{a}_1, \mathbf{a}_2, \mathbf{a}_3$ are the lattice vectors. In equation (2) we have also introduced a shape function $s(\mathbf{r})$ equal to the unity inside the volume $V$ of the crystal and zero outside (so-called Ewald function [39]) where $s(\mathbf{r})$ stands for the finite size of the sample

$$s(\mathbf{r}) = \begin{cases} 1 & \text{for} \quad \mathbf{r} \in V \\ 0 & \text{for} \quad \mathbf{r} \notin V \end{cases} \quad . \tag{3}$$

---

[1] Here we assume that the kinematical approximation for the description of X-ray scattering on crystalline samples is valid. This is a good approximation for scattering of X-rays in the range of 10 keV and submicron crystal sizes. However, if crystalline particles reach few micron size, multiple scattering, or dynamical effects [35,36] could become important [37]. For these crystal sizes refraction effects should be also considered [38].

The shape function $s(\mathbf{r})$ leads to a "spreading" of the $\delta$-type intensity distributions around the Bragg peaks which characterize an infinite crystal. The scattering amplitude $A(\mathbf{q})$ in equation (1) can now be conveniently calculated by the means of the approach originally proposed by von Laue [40], which reduces the sum over the points of the ideal lattice within the volume of the finite crystal to an integral over all space.

Substituting now expression (2) for the electron density into (1) and using convolution theorem we get for the scattered amplitude

$$A(\mathbf{q}) = F(\mathbf{q}) \cdot \rho_\infty(\mathbf{q}) \otimes s(\mathbf{q}) \tag{4}$$

Here

$$F(\mathbf{q}) = \int \rho_{uc}(\mathbf{r}) e^{-i\mathbf{q}\cdot\mathbf{r}} d^3r = \sum_j f_j(\mathbf{q}) e^{-i\mathbf{q}\cdot\mathbf{r}_j} \tag{5}$$

is the structure factor of the unit cell and

$$f_j(\mathbf{q}) = \int \rho_j(\mathbf{r}) e^{-i\mathbf{q}\cdot\mathbf{r}} d^3r \tag{6}$$

is the atomic scattering factor of the atom $j$ in the unit cell and integration is performed over the volume of the unit cell. It is also assumed that the structure factors of the different cells are identical, as it is in general for perfect crystals. Usually, the structure factor $F(\mathbf{q})$ is a complex function. In equation (4)

$$s(\mathbf{q}) = \int s(\mathbf{r}) e^{-i\mathbf{q}\cdot\mathbf{r}} d^3r \tag{7}$$

is the Fourier transform of the shape function $s(\mathbf{r})$ and $\rho_\infty(\mathbf{q})$ is the Fourier transform of the lattice function that reduces to the sum of $\delta$-functions

$$\rho_\infty(\mathbf{q}) = \int \rho_\infty(\mathbf{r}) e^{-i\mathbf{q}\cdot\mathbf{r}} d^3r = \frac{(2\pi)^3}{v} \sum_n \delta(\mathbf{q}-\mathbf{h}_n) \tag{8}$$

where $v$ is the volume of the unit cell, $\mathbf{h}_n = h\mathbf{h}_1 + k\mathbf{h}_2 + l\mathbf{h}_3$ being the reciprocal lattice vectors and the summation is carried out over all reciprocal lattice points. In summary, we obtain for the scattered amplitude (4)

$$A(\mathbf{q}) = \frac{F(\mathbf{q})}{v} \sum_n A_n(\mathbf{q}-\mathbf{h}_n) \tag{9}$$

with $A_n(\mathbf{q}-\mathbf{h}_n) = s(\mathbf{q}-\mathbf{h}_n)$ being the amplitude scattered in the vicinity of the reciprocal lattice vector $\mathbf{h}_n$. From this expression we can see that the scattering amplitude is directly connected with the FT of the 'shape' function $s(\mathbf{r})$. Here it is also important to note that the structure factor $F(\mathbf{q})$ is extended in reciprocal space over many reciprocal lattice points. In the limit of the infinite crystal FT of the shape function $s(\mathbf{q})$ (7) reduces to the $\delta$-function and we get for the amplitude (9)

$$A(\mathbf{q}) = \frac{F(\mathbf{q})}{v} \sum_n \delta(\mathbf{q} - \mathbf{h}_n) \qquad (10)$$

It is a well-known result in crystallography that the scattering amplitude of the infinite periodic object is sampled at its reciprocal lattice or Bragg points and, in principle, no information is available between these sampling points. Taking the inverse FT of (10) we obtain a well known crystallographic formula for the electron density of the unit cell expressed through the Fourier components of structure factors

$$\rho(\mathbf{r}) = \frac{1}{v} \sum_{\mathbf{h}_n} F(\mathbf{h}_n) e^{i\mathbf{h}_n \cdot \mathbf{r}} \qquad (11)$$

Unfortunately, only the amplitudes $|F(\mathbf{h}_n)|$ of the structure factors can be measured in experiment and there is no direct phase information available (so-called phase problem in crystallography).

For a crystal of macroscopic dimensions $D \gg a$, where $a$ is the size of a unit cell, the function $s(\mathbf{q})$ has appreciable values only for $\Delta q \sim 2\pi/D$ much smaller than the reciprocal lattice parameters $h_n = (2\pi/a)n$. Thus according to (9) and neglecting the small cross terms, the intensity scattered by the crystal of finite dimensions will be determined by the sum over reciprocal lattice points

$$I(\mathbf{q}) = |A(\mathbf{q})|^2 = \frac{|F(\mathbf{q})|^2}{v^2} \sum_n |A_n(\mathbf{q} - \mathbf{h}_n)|^2 \qquad (12)$$

In the vicinity of the reciprocal point $\mathbf{h}_n = \mathbf{h}$, $\mathbf{q} \approx \mathbf{h}$, the intensity distribution can be written as

$$I_h(\mathbf{Q}) = \frac{|F(\mathbf{h})|^2}{v^2}|A_h(\mathbf{Q})|^2 \quad (13)$$

where $\mathbf{Q} = \mathbf{q} - \mathbf{h}$ and $A_h(\mathbf{Q}) = s_h(\mathbf{Q})$.

In the case of the infinite crystal we have from (10)

$$I(\mathbf{q}) = |A(\mathbf{q})|^2 = \frac{|F(\mathbf{q})|^2}{v^2}\sum_n \delta(\mathbf{q} - \mathbf{h}_n) \quad (14)$$

According to this mathematics, several important points have to be outlined. First of all, according to (10) and (14) in the case of the infinite crystal the scattering amplitude (or structure factor of the unit cell), and thus the intensity, is sampled at fixed points in reciprocal space. These points correspond to the nodes of the reciprocal lattice $\mathbf{h}_n$. Generally, no experimental information is available in between these sampling points and there is no way to measure continuous diffraction patterns from infinite crystal. Hence, there is no possibility to oversample diffraction patterns from infinite crystals. This was noted first by D. Sayre [41] in early 50-s. It was also noted by the same author that sampling of the reciprocal space at Bragg points corresponds exactly to Nyquist sampling of the electron density of the unit cell of size $a$ (for simplicity we will limit our discussion with 1D case). In other words, if we have an infinite periodic density $\rho(x)$ with the unit cell size $a$ (see Fig. 3), the corresponding reciprocal space amplitude $F(q)$ will be sampled at Bragg points $h_n = (2\pi/a)n$. At the same time according to the Shannon theorem [16] the minimum sampling distance for the electron density of a unit cell with size $a$ will be $\Delta q = 2\pi/a$, which is identical to the sampling at the Bragg positions. According to the same theorem there is no additional information between these sampling points. If all values of the amplitude

$F(q)$ at the sampling points are known then a continuous amplitude could be constructed *uniquely* as

$$F(q) = \frac{2\pi}{a}\sum_{h_n} F(h_n)\frac{\sin[a/2(q-h_n)]}{\pi(q-h_n)}. \quad (15)$$

If the complex structure factors $F(h_n)$ (including the phases) could be measured it would be possible to construct a continuous function $F(q)$ according to 15). Then, applying the inverse FT, the electron density in the unit cell could be reconstructed. However, as mentioned above the phases of $F(h_n)$ are missing in the measurements. So, in the case of the infinite

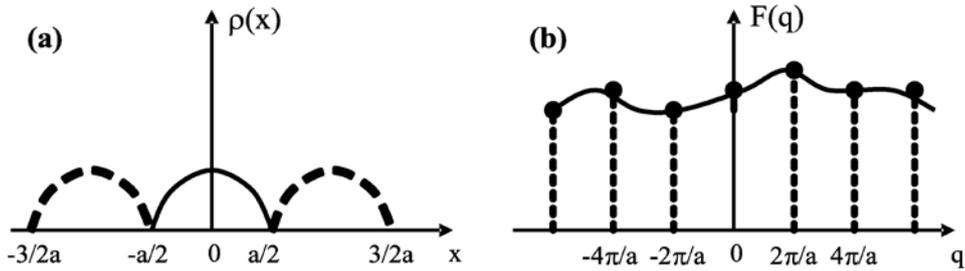

**Fig. 3** (a) Periodic electron density $\rho(x)$ with the unit cell size $a$. (b) Due to a periodicity of the electron density structure factor $F(q)$ is sampled at Bragg points $h_n = (2\pi/a)n$. No additional data points can be measured in between these sampling points, so in traditional crystallography oversampling is not possible.

periodic sample it appears impossible to obtain a continuous diffraction pattern and consequently to apply an oversampling method for phase retrieval methods. Situation looks even worse when the intensity distribution $I(\mathbf{q})$ (14) is analyzed: The intensity distribution can be presented as a FT of the autocorrelation function of the electron density. For the electron density of the unit cell of size $a$, the corresponding autocorrelation function would have an extension of $2a$ with corresponding minimum Nyquist sampling frequency $\Delta q = \pi/a$. That means that we cannot make use of the sampling theorem (15) to reconstruct continuous intensity distribution function $I(\mathbf{q})$, because just half of the necessary data points are missing. As a consequence phase problem in crystallography is *twice* underdetermined. That makes phase retrieval for crystallography even more difficult problem then in optics.

The situation is quite different if, instead of an infinite crystal, a crystal of finite size is illuminated with coherent beams. Then the intensity distribution is given by (12) and is therefore the FT of a shape function $s(\mathbf{r})$ around *each* of the Bragg points. It is clear that in this case the intensity distribution around each Bragg point will be a continuous function and can be, in principle, oversampled to necessary level. Hence, a unique reconstruction of the crystal shape is possible.

Some general properties of this intensity distribution have to be outlined. For an arbitrary shap of the unstrained crystal, the intensity distribution (12) is a periodic function of $\mathbf{q}$ (see Fig. 4). The intensity distribution is locally centrosymmetric around *each* $\mathbf{h}_n$ and has the same shape for *each* reciprocal lattice point $\mathbf{h}_n$. It takes its maximum value of $|F(\mathbf{h}_n)|^2 V^2/v^2$, where $V$ is the volume of the crystal, if the scattering vector is exactly equal to $\mathbf{q} = \mathbf{h}_n$ and this point is the center of symmetry of the intensity distribution $I(\mathbf{q})$ (since according to (7) $s(-\mathbf{q}) = s^*(\mathbf{q})$). As follows from equation (12) the simplest picture of identical repeated distributions arises in unstrained crystals of any arbitrary shape. The detailed 3D shape of this distribution is determined by the FT of the crystal shape function $s(\mathbf{r})$ (see Eq.(7)). The intensity distribution measured by the 2D detector depends also on the Bragg angle and on the deviation from the exact Bragg conditions (the detector plane is always perpendicular to $\mathbf{k}_f$ vector). If the $\mathbf{z}$-axis in reciprocal space is directed along the $\mathbf{k}_f$ vector and the detector is positioned at a Bragg angle, then we have from Eqs. (7) to (13) the following distribution of the amplitude in reciprocal space

$$A(Q_x, Q_y) = \frac{F(\mathbf{h})}{v} \int s_z(x, y) e^{-iQ_x x - iQ_y y} dxdy \qquad (16)$$

where $s_z(x,y) = \int s(x,y,z)dz$ is the *projection* of the crystal shape on the $(x,y)$-plane that is defined as perpendicular to the $\mathbf{k}_f$ vector. Obviously, the inverse FT of the amplitude distribution $A(Q_x, Q_y)$ will recover this projection of the crystal shape. It is also clear that for a general crystalline sample this projection is equivalent to the projection of the *electron density* of a sample to the same plane.

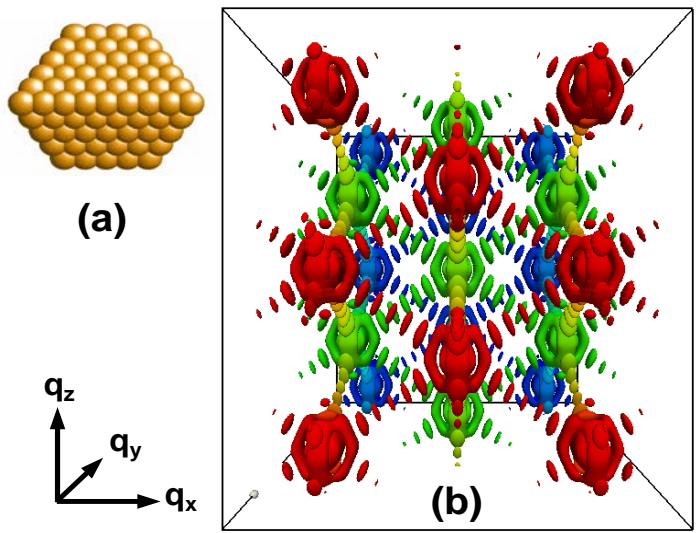

**Fig. 4** A finite size crystal (a) and its representation in reciprocal space (b). Different color corresponds to a different location of Bragg peaks in $q_y$ direction.

In Fig. 5 diffraction patterns calculated for different crystal shapes and described scattering conditions are presented. The cut of the reciprocal space for other values of $Q_z$ than $Q_z = 0$ can also be calculated using Eq. (16). Simply, it requires the replacement of the real function $s_z(x,y)$ by a complex valued function $s_z(x,y) = \int s(x,y,z) \exp(iQ_z z)dz$. Examples of such calculations are presented in Fig. 6 for different values of $Q_z$ and the crystal shape shown in the left panel of Fig. 5. We want to note here that due to the presence of the phase factor $\exp(iQ_z z)$ each cut of the reciprocal space by the Ewald sphere at $Q_z \neq 0$ will produce non-centrosymmetric intensity distribution even for an unstrained crystal (see Fig. 6). However, the 3D intensity distribution will be centrosymmetric around each Bragg point (see Fig. 4).

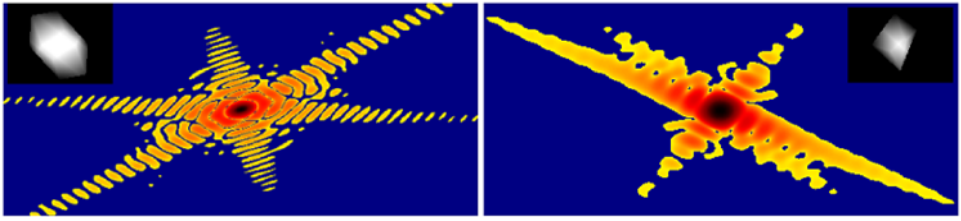

**Fig. 5** Projection of the different crystal shapes on the plane perpendicular to $\mathbf{k}_f$ and corresponding diffraction pattern calculated at exact the Bragg position. Adapted from Ref. [29].

In principle, the whole 3D intensity distribution $I(\mathbf{Q})$ can be measured by a 2D detector and by scanning the sample near Bragg position or changing the incident energy. This 3D distribution can also be directly inverted giving the shape of the crystalline part of the sample in 3D (see for the first demonstration [42]). Remarkably, in the case of the 3D phase retrieval the necessary conditions on oversampling in the third dimension are quite relaxed as was noted by Millane [43]. In practice, few tens of angular scans in reciprocal space are sufficient to invert the 3D diffraction pattern of a micron size crystalline particle (see for details [42,44]).

As was proposed by von Laue [40] the Green's theorem can be applied to Eq. (7) and the volume integral can be transformed to an integral over the external surface area ($S$) of the crystal

$$s(\mathbf{q}) = \frac{i}{q^2} \int_S (\mathbf{q} \cdot \mathbf{n}) e^{-i\mathbf{q} \cdot \mathbf{r}} d\sigma, \qquad (17)$$

where the unit vector $\mathbf{n}$ is an outward normal to the crystal. The maximum of this distribution for the flat surface is along directions normal to the surface. The existence of these flares was predicted by von Laue [40] and given the name "Stacheln" (spike). They have been experimentally observed later in the studies of the surface diffraction in the form of crystal truncation rods (CTR) [45], or asymptotic Bragg diffraction [46]. In the case of a crystal with a center of symmetry and with a pair of identical opposite facets we obtain from (17)

$$s(\mathbf{q}) = 2\frac{(\mathbf{q}\cdot\mathbf{n})}{q^2}\int_S \sin(\mathbf{q}\cdot\mathbf{r})d\sigma \quad (18)$$

If the distance between the facets is equal to $D$, then for the direction of $\mathbf{q}$ perpendicular to the facets we finally get

$$s(q) = \frac{2}{q} S \sin(qD/2) \quad (19)$$

It follows immediately, that for two opposite facets and coherent illumination *interference* patterns appear in the intensity distribution rather than a smooth $q^{-2}$ decrease of intensity for a single surface (see Fig. 5).

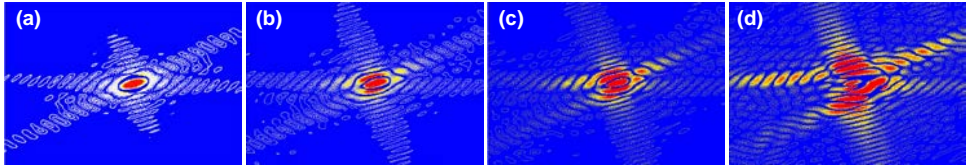

**Fig. 6** The cross section of the reciprocal space of the diffraction pattern calculated from the crystal shape shown in the inset of Fig. 12.5 (left panel) for different $q_z$ values: (a) $q_z = 0$, (b) $q_z = 0.357 q_D$, (c) $q_z = 0.476 q_D$ and (d) $q_z = 1.19 q_D$. Here $q_D$ correspond to the fringe spacing $q_D \sim 2\pi/D$. Intensity in figure is rescaled for clarity. Adapted from Ref. [29].

This interference pattern is similar to the fringes from slit scattering in optics, when the slit is illuminated by a coherent light. The integral width of this intensity distribution in reciprocal space is equal to $\delta q = 2\pi/D$. This leads to a rod-like shape intensity distribution for crystals shaped like a compressed disc. Such behavior was, for example, observed in the study of thin films of $Cu_3Au$ [47]. In the case of a flat surface the same Green's theorem can be applied once more to equation (17) transforming the surface integral to an integral around the boundary of the facet $S$. Now, diffraction from the edges will produce, instead of truncation rods, crystal truncation planes (CTP) (see for the first observation of CTP [48]).

As a conclusion we can summarize that any pair of opposite facets and corresponding edges of unstrained crystals in a coherent beam produces an interference pattern with the maximum intensity distribution along the normal to the facet (CTR), or perpendicular to the opposite surface edges (CTP).

Equation (12) corresponds to the situation when *one* particle is illuminated by a coherent beam. In the case when two or more crystallites are located at some distance from each other and are illuminated by the same coherent beam additional interference terms will appear in the expression for the intensity (12). Especially interesting (but not discussed here) is the case when a small particle is well separated from a big one and illuminated by the same coherent beam. This is similar to the principles of Fourier holography, when the object can be found as one term in the autocorrelation [49-51].

Up to now we presented the mathematics of diffraction patterns which are measured locally around a fixed Bragg point for a crystal of the finite size. However, another kind of measurement can be suggested, when diffraction patterns are measured simultaneously around several reciprocal lattice points. Taking into account that this distribution is measured for a finite number of reciprocal lattice points up to $q \sim q_{max}$ the complex amplitude distribution can be written as

$$A(\mathbf{q}) = B(\mathbf{q}) \cdot F(\mathbf{q}) \cdot [\rho_\infty(\mathbf{q}) \otimes s(\mathbf{q})], \quad (20)$$

where $B(\mathbf{q})$ is an envelope function with the effective size $q \sim q_{max}$. Inverting this relationship with the inverse FT yields in real space an electron density

$$\rho(\mathbf{r}) = [b(\mathbf{r}) \otimes \rho_{uc}(\mathbf{r})] \otimes [\rho_\infty(\mathbf{r}) \cdot s(\mathbf{r})]$$ , (21)

where $b(\mathbf{r})$ is the inverse Fourier transform of $B(\mathbf{q})$. This electron density is peaking at the regular positions of the unit cell due to the function $\rho_\infty(\mathbf{r})$ and has an overall shape of the sample $s(\mathbf{r})$. Most importantly, it contains the position of the atoms in the unit cell due to the reconstruction of the electron density function of a unit cell $\rho_{uc}(\mathbf{r})$. This means the following: If the continuous intensity distribution around several Bragg peaks will be measured simultaneously and phase retrieval methods will be applied to get the phase, then, in principle, the electron density with *atomic* resolution will be obtained. The resolution in real space around each atomic position in such experiments will be determined by the area accessed in the measurements in reciprocal space and can be estimated to be about $\Delta r \sim 2\pi / q_{max}$.

In order to map several Bragg peaks simultaneously with one detector different approaches can be applied. For conventional crystalline samples with a unit cell of the size of a few angstroms hard X-rays in the range of 100 keV should be used [52]. Another approach is to use long period crystalline samples such as colloidal crystals with a typical unit cell sizes on the order of a few hundred nanometers. The latter was successfully realized in CXDI experiments on colloidal 2D and 3D samples [53,54] (see section 3.1 ). High energy electron beams in nano-diffraction experiments with the conventional transmission electron microscope (TEM) can be also used for realization of these ideas [55-57].

### *2.2 Coherent scattering from a finite size crystal with a strain*

In the case of a deformed crystal we can write the electron density as the sum of terms corresponding to individual atoms

$$\rho(\mathbf{r}) = \sum_{n=1}^{N}\sum_{j=1}^{S} \rho_{nj}(\mathbf{r} - \mathbf{R}_{nj} - \mathbf{u}(\mathbf{R}_{nj})) \qquad (22)$$

where $\mathbf{R}_{nj} = \mathbf{R}_n + \mathbf{r}_j$ and $\mathbf{u}(\mathbf{R}_{nj})$ is the displacement from the ideal lattice point. Summation in (22) is performed over $N$ unit cells which contain $S$ atoms. We assume here that the whole crystal is coherently illuminated. Substituting this expression for the electron density into (1) and changing variables in each term, the scattering amplitude can be written as a sum over the unit cells

$$A(\mathbf{q}) = \sum_{n=1}^{N} F_n(\mathbf{q}) e^{-i\mathbf{q}\cdot\mathbf{u}(\mathbf{R}_n)} e^{-i\mathbf{q}\cdot\mathbf{R}_n} \qquad (23)$$

where $F_n(\mathbf{q})$ is the complex valued structure amplitude of the $n^{th}$ cell. Here it is assumed that all atoms in the unit cell are displaced uniformly $\mathbf{u}(\mathbf{R}_{nj}) \equiv \mathbf{u}(\mathbf{R}_n + \mathbf{r}_j) = \mathbf{u}(\mathbf{R}_n)$. It is important to note that equation (23) is also valid for the more general case allowing different displacements of atoms in different unit cells but with another definition of the structure amplitude $F_n(\mathbf{q})$ [58].

Now we will consider, as before, the scattering of X-rays on a crystal with finite size. The scattering amplitude $A(\mathbf{q})$ (23) for the crystal of finite dimensions will be calculated using the same approach as described in the previous section. According to this approach equation (23) can be identically rewritten in the form

$$A(\mathbf{q}) = F(\mathbf{q}) \int \rho_{\infty}(\mathbf{r}) S(\mathbf{r}) e^{-i\mathbf{q}\cdot\mathbf{r}} d^3 r \qquad (24)$$

where it is assumed that the structure factors of the different cells are identical with $F_n(\mathbf{q}) = F(\mathbf{q})$ and integration is carried out over the whole space. In this equation we have introduced a *complex* function

$$S(\mathbf{r}) = s(\mathbf{r})\exp(-i\mathbf{q}\cdot\mathbf{u}(\mathbf{r})) \qquad (25)$$

with the shape function $s(\mathbf{r})$ (3) as an amplitude and the phase $\phi(\mathbf{r}) = \mathbf{q}\cdot\mathbf{u}(\mathbf{r})$, where the deformation field $\mathbf{u}(\mathbf{r})$ is included. It is important to note that no restrictions on the shape of the crystal and the deformation field apply.

Performing now the same calculations as in the previous section we will obtain for the amplitude (see Eq. (4))

$$A(\mathbf{q}) = F(\mathbf{q})\cdot S(\mathbf{q}) \otimes \rho_\infty(\mathbf{q}), \qquad (26)$$

where $S(\mathbf{q})$ is the Fourier integral of $S(\mathbf{r})$

$$S(\mathbf{q}) = \int S(\mathbf{r})e^{-i\mathbf{q}\cdot\mathbf{r}}d^3r, \qquad (27)$$

and the integration over $d^3r$ is carried out over the whole space.

Now using the expression for the FT of $\rho_\infty(\mathbf{r})$ (8) we can derive the amplitude (26)

$$A(\mathbf{q}) = \frac{F(\mathbf{q})}{v}\sum_n A_n(\mathbf{q}-\mathbf{h}_n), \qquad (28)$$

with $A_n(\mathbf{q}-\mathbf{h}_n) = S(\mathbf{q}-\mathbf{h}_n)$, now containing the deformation values. From this expression we can see that the scattering amplitude is directly connected with the FT of the complex 'shape' function $S(\mathbf{r})$ and its phase for the fixed reciprocal lattice point $\mathbf{h}$ is a sum of phases of the structure factor $F(\mathbf{h})$ and function $S(\mathbf{q})$.

For a crystal of macroscopic dimensions we will again get a periodic intensity distribution

$$I(\mathbf{q}) = |A(\mathbf{q})|^2 = \frac{F(\mathbf{q})}{v}\sum_n |A_n(\mathbf{q}-\mathbf{h}_n)|^2, \qquad (29)$$

where each term gives intensity values close to reciprocal point,

$$I_h(\mathbf{Q}) = \frac{|F(\mathbf{h})|^2}{v^2} |A_\mathbf{h}(\mathbf{Q})|^2 \quad (30)$$

Here the amplitudes $A_\mathbf{h}(\mathbf{Q})$ are defined as

$$A_\mathbf{h}(\mathbf{Q}) = \int s(\mathbf{r}) e^{-i\mathbf{h}\cdot\mathbf{u}(\mathbf{r})} e^{-i\mathbf{Q}\cdot\mathbf{r}} d^3 r, \quad (31)$$

and as before $\mathbf{Q} = \mathbf{q} - \mathbf{h}$.

In the following we will outline the differences between the coherent diffraction pattern of the unstrained crystal and the crystal with the deformation field. In this case for any arbitrary form of the crystal and the strain field, the intensity distribution (29) is still localized around reciprocal lattice vectors $\mathbf{h}_n$. However, in contrast to the unstrained case, for samples with arbitrary strain the intensity distribution *locally* is not centrosymmetric around $\mathbf{h}_n$ and the shape differs at every reciprocal lattice point $\mathbf{h}_n$. Effects associated with the strain $\mathbf{u}(\mathbf{r})$ lead to different distributions near different reciprocal-lattice points.

This is illustrated in Fig. 7, where a two-dimensional example of the diffraction from a strained object is shown. In Fig. 7(a) a real envelope function $s(\mathbf{r})$, defining a crystal shape is shown. It is obtained from a scanning-electron microscope image of a partially annealed array of gold nanocrystals on a glass substrate. In Fig. 7(b) its Fourier transform, which shows the expected local symmetry is depicted. Fig. 7(c) is a representation of a complex envelope function $S(\mathbf{r})$, where the amplitude is the same as before and the phase varies quadratically as a function of the radius from the object's center. The shaped (gray) rings represent phase reversals. The FT (Fig. 7(d)) shows the diffraction pattern expected from such a strained particle. Clearly, the local symmetry of the unstrained case on the left is distorted but still the main features, the flares and the fringes are resembled. This example shows the main point of our discussion, that the *shape* of a small crystal is related to the symmetric part of its coherent diffraction pattern, while its internal *strain* appears as an asymmetry in the diffraction.

Examples of successful experimental realization of these ideas in mapping the strain field in thin crystalline layers, or in isolated crystalline particles can be found in [44,60-63] (see also for the review [8]) .

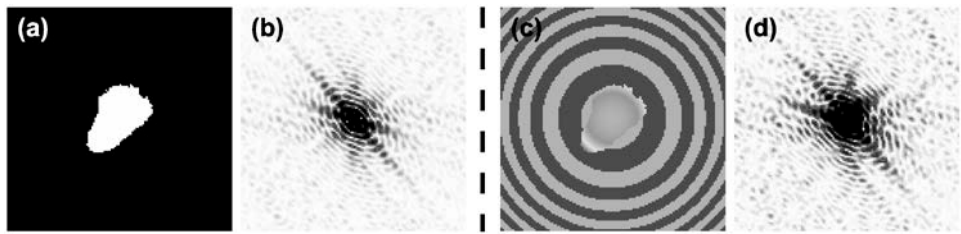

**Fig. 7** Illustration of the effects of strain on the diffraction from a 2D crystal. (a) is an unstrained object and (b) - its calculated coherent diffraction pattern. (c) is the object with the addition of a real space strain increasing quadratically with radius from the center of the object and (d) - its diffraction pattern. The alternate circular shading denotes positions for which the phase lies between 0 and $\pi$ or between $\pi$ and $2\pi$. Adapted from Ref. [59]

## *2.3 Partially coherent scattering from a finite size crystal*

In the previous section the case of totally coherent incident radiation was considered. Now, we will assume that the incoming beam is partially coherent. The general properties of partially coherent radiation are discussed in detail in a number of textbooks [64-66]. We will reformulate the general results for the special case of scattering of partially coherent X-ray radiation on small crystalline particles (see also [29,30,67], for the case of partially coherent X-ray scattering on surfaces see [68,69] and for CXDI with the partially coherent X-rays in the forward direction [31]). The whole problem separates into two parts. For the first part we consider scattering of radiation with an arbitrary state of coherence from a small crystalline particle. For the second part a special realization of the incoherent source with a Gaussian intensity distribution will be discussed.

Let us assume the incident radiation to be a quasi-monochromatic wave with one polarization state of the electric field,

$$E_{in}(\mathbf{r},t) = A_{in}(\mathbf{r},t)e^{i\mathbf{k}_i\cdot\mathbf{r}-i\omega t}, \qquad (32)$$

where $|\mathbf{k}| = 2\pi/\lambda$. Here $\lambda$ and $\omega$ are the average wavelength and frequency of the beam. The amplitude $A_{in}(\mathbf{r},t)$ is a slowly varying function with spatial variations much bigger than the wavelength $\lambda$ and time scales much larger than $1/\omega$. Consequently, according to the standard Huygens-Fresnel principle [64] in the limits of kinematical scattering, the amplitude of the wavefield $E_{out}(\mathbf{v},t)$, after being scattered from the sample to position $\mathbf{v}$ at the detector (see Fig. 8), can be written as[2]

$$E_{out}(\mathbf{v},t) = \int \rho(\mathbf{r}) \frac{A_{in}(\mathbf{r},t-\tau_r)}{l_r} e^{i\mathbf{k}_i\cdot\mathbf{r}-\omega(t-\tau_r)} d^3r, \qquad (33)$$

where $l_r$ is the distance between points $\mathbf{r}$ and $\mathbf{v}$ with the origins in the sample center and detector plane respectively, $\tau_r = l_r/c$ is the time delay for the radiation propagation between these points and $c$ is the speed of light. In this expression for the scattered wavefield we have neglected the absorption as the small sample is small and set the obliquity factor to unity.

---

[2] In this expression and below we will omit all not essential integral prefactors.

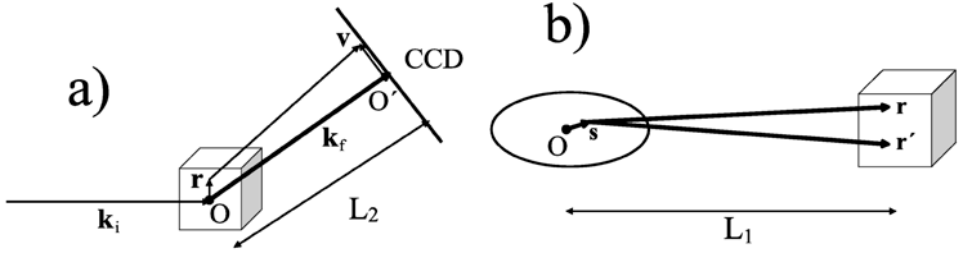

**Fig. 8** Definition of notations used for the scattering geometry in calculation of scattering of partial coherent radiation. a) X-ray beam is scattered on a small crystal particle and intensity is measured at the distance $L_2$ from the sample by 2D detector (for example, CCD detector). b) Synchrotron source produce incoherent beam on the distance $L_1$ from the sample. Adapted from Ref. [29].

We will define the amplitude of the scattered wave $A(\mathbf{v},t)$ in the usual way by $E_{out}(\mathbf{v},t) = A(\mathbf{v},t)(e^{ikL_2}/L_2)e^{i\omega t}$. Obviously, we can assume that the distance from the object to detector is $L_2 \gg D$, where $D$ is a typical size of an object. Then, in the limits of paraxial approximation for the distance $l_r$ between points $\mathbf{r}$ and $\mathbf{v}$, we can use an expansion $l_r \approx L_2 - \mathbf{n}_f \cdot \mathbf{r} + (\mathbf{v}-\mathbf{r})^2/(2L_2)$, where $\mathbf{n}_f = \mathbf{k}_f/|\mathbf{k}_f|$. Here, we also assumed that the detector plane is perpendicular to $\mathbf{k}_f$. Substituting this expansion into equation (33) we obtain for the scattering amplitude

$$A(\mathbf{v},t) = \int \rho(\mathbf{r}) A_{in}(\mathbf{r},t-\tau_r) P_{L_2}(\mathbf{v}-\mathbf{r}) e^{-i\mathbf{q}\cdot\mathbf{r}} d^3r \tag{34}$$

where $P_{L_2}(\mathbf{v}-\mathbf{r})$ is the Green's function (or propagator), that describes the propagation of radiation in free space. In the frame of the same theory this function is equal to

$$P_{L_2}(\mathbf{v}-\mathbf{r}) = \frac{1}{i\lambda L_2} e^{i(k/2L_2)(\mathbf{v}-\mathbf{r})^2} \tag{35}$$

We are interested in the far-field (or Fraunhofer) limit of equation (34), when condition $kD^2/(2L_2) \ll 1$ is satisfied[3]. For the typical X-ray coherent

---

[3] Near-field Fresnel conditions can be also considered in the frame of the same theory. This will bring just to appearance of the additional phase factor $\exp[i(k/2L_2)\mathbf{r}^2]$ in

experiment with radiation energy $E_\gamma \approx 8$ keV and detector at a distance $L_2 \approx 3$ m, this condition limits the size of the particles to $D \ll 10\,\mu$m. In this limit we can neglect the $(k/2L_2)\mathbf{r}^2$ term in the exponent (35) and have for the propagator

$$\lim_{(kD^2)/L_2 \to 0} P_{L_2}(\mathbf{v}-\mathbf{r}) \to \frac{1}{i\lambda L_2} \exp[i(k/2L_2)\mathbf{v}^2]\exp(-\mathbf{q}_v \cdot \mathbf{r}) ,$$

where $\mathbf{q}_v = (k/L2)\mathbf{v}$. So, in the far-field we obtain for the amplitude (34)

$$A(\mathbf{q}',t) = \int \rho(\mathbf{r}) A_{in}(\mathbf{r}, t-\tau_r) e^{-i\mathbf{q}'\cdot\mathbf{r}} d^3r \tag{36}$$

where $\mathbf{q}' = \mathbf{q} + \mathbf{q}_v$. Here, we have omited the phase term $\exp[i(k/2L_2)\mathbf{v}^2]$ before the integral as it will cancel while calculating intensities at the same point $\mathbf{v}$ at the detector plane. We would like to note here, that this expression coincides with the coherent amplitude of equation (1) in the limit $A_{in}(\mathbf{r}, t-\tau_r) \to 1$. Now, we will consider the case when the scattering particle is a crystalline sample with a periodic electron density function and the amplitude $A_{in}(\mathbf{r}, t-\tau_r)$ is a slow varying function on the size of the unit cell. Under this considerations performing the same transformations as in the previous section we finally obtain

$$A(\mathbf{q}',t) = \frac{F(\mathbf{q})}{v} \sum_n A_n(\mathbf{q}'-\mathbf{h}_n, t) \tag{37}$$

where

$$A_n(\mathbf{q},t) = \int s(\mathbf{r}) A_{in}(\mathbf{r}, t-\tau_r) e^{-i\mathbf{q}\cdot\mathbf{r}} d^3r \tag{38}$$

is a scattering amplitude in the direction of reciprocal vector $\mathbf{h}_n$. Here, as before, $s(\mathbf{r})$ is a shape function of the crystal and for simplicity we assume that crystal is unstrained. However this result can be generalized also for the integrand of Eq. (12.36) (see for details [30]).

the case of the strained crystal by adding the exponential factor $\exp[-i\mathbf{q}\cdot\mathbf{u}(\mathbf{r})]$ to the integral (38).

According to (37) and (38), the intensity of the scattered radiation measured at the position $\mathbf{v}$ of the detector near one of the Bragg points $\mathbf{h}_n = \mathbf{h}$ (we denote the corresponding amplitude as $A_h(\mathbf{q},t)$) is equal to

$$I(\mathbf{Q}) = <A(\mathbf{Q},t)A^*(\mathbf{Q},t)>_T = \frac{|F(\mathbf{h})|^2}{v^2} <|A_h(\mathbf{Q},t)|^2>_T \qquad (39)$$
$$= \frac{|F(\mathbf{h})|^2}{v^2} \iint s(\mathbf{r})s(\mathbf{r'})\Gamma_{in}(\mathbf{r},\mathbf{r'},\Delta\tau)e^{-i\mathbf{Q}\cdot(\mathbf{r}-\mathbf{r'})}d^3r d^3r',$$

where $\mathbf{Q} = \mathbf{q'} - \mathbf{h} = \mathbf{q}_v + \mathbf{q} - \mathbf{h}$, $\Delta\tau = (l_r - l_{r'})/c$ is a time delay and

$$\Gamma_{in}(\mathbf{r},\mathbf{r'},\tau) = <A_{in}(\mathbf{r},t)A_{in}^*(\mathbf{r'},t+\tau)>_T \qquad (40)$$

is the mutual coherence function. The averaging $<>_T$ in (39) and (40) is conducted for a time $T$ which is much longer then the time of fluctuation of the X-ray field, and it is assumed that the incoming radiation is ergodic and stationary.

For the case of cross-spectral pure light we can write the mutual coherence function as a product [65,66]

$$\Gamma_{in}(\mathbf{r},\mathbf{r'},\tau) = \sqrt{I(\mathbf{r})}\sqrt{I(\mathbf{r'})}\gamma_{in}(\mathbf{r}.\mathbf{r'})F(\tau) \qquad (41)$$

Here $I(\mathbf{r}) = <|A_{in}(\mathbf{r},t)|^2>_T$ and $I(\mathbf{r'}) = <|A_{in}(\mathbf{r'},t)|^2>_T$ are the averaged intensities of the incoming radiation at points $\mathbf{r}$ and $\mathbf{r'}$, $\gamma_{in}(\mathbf{r},\mathbf{r'})$ is the complex degree of coherence and $F(\tau)$ is the time autocorrelation function.

We will make some more simplifying considerations to have an explicit form for the mutual coherence function (41). We will assume that the incident radiation is coming from a planar incoherent source with Gaussian distribution of intensity which is located at a distance $L_1$ from the sample (Fig. 8b). This will be an approximation for the actual 3D X-ray source from the synchrotron storage ring. We will also consider that the distance $L_1$ is much larger than the size of the particle $D$ and, than the average size of the source $S$. According to the van Cittert-Zernike theorem [65,66], the complex degree of coherence can be obtained in the same limit of paraxial approximation (see [70] for the generalization of this approach)

$$\gamma_{in}(\mathbf{r},\mathbf{r}') = \frac{e^{i\psi}\int I(\mathbf{s})e^{i\mathbf{s}\cdot(\mathbf{r}-\mathbf{r}')k/L_1}d^3s}{\int I(\mathbf{s})d^3s} \quad (42)$$

where the phase factor is $\psi = (k/2L_1)(r^2 - r'^2)$, $I(\mathbf{s})$ is the intensity distribution of the incoherent source and integration is performed over the whole area of the incoherent source. It is interesting to note here that for an incoherent source expression (42) is exact up to second order terms in $\mathbf{s}$. For the typical CXD experiment at a synchrotron source with a distance from source to sample $L_1 \approx 40$ m and an energy of $E_\gamma \approx 8$ keV the far-field conditions $kD^2/(2L_1) << 1$ can easily be satisfied, giving the upper limit for the size of the particle as $D << 40$ μm. In this far-field limit we can neglect the phase prefactor $e^{i\psi}$ in Eq. (42) and consider the complex degree of coherence $\gamma_{in}(\mathbf{r},\mathbf{r}')$ (42) as a real valued function. Following this model of an incoherent source in the far-field limit the intensity of the incoming radiation at points $\mathbf{r}$ and $\mathbf{r}'$ at the sample can be calculated as: $I(\mathbf{r}) \approx I(\mathbf{r}') = I_0 = (\lambda/L_1)^2 \int I(\mathbf{s})d^2s$.

It is common to describe the intensity distribution of the synchrotron source by the Gaussian function

$$I(s_x, s_y) = \frac{I_0}{2\pi\sigma_x\sigma_y} e^{-(s_x^2/\sigma_x^2 + s_y^2/\sigma_y^2)/2} \quad (43)$$

where $\sigma_x$ and $\sigma_y$ are the halfwidths of the intensity distribution in $x$ and $y$ directions. Due to the fact that (42) is a FT we get also a Gaussian form for the complex degree of coherence

$$\gamma_{in}(\mathbf{r}_\perp - \mathbf{r}'_\perp) = \exp\left(-\frac{(\mathbf{r}_\perp - \mathbf{r}'_\perp)^2}{2\xi_\perp^2}\right) = \exp\left(-\frac{(x-x')^2}{2\xi_x^2} - \frac{(y-y')^2}{2\xi_y^2}\right) \qquad (44)$$

Here $\mathbf{r}_\perp$ and $\mathbf{r}'_\perp$ are projections of $\mathbf{r}$ and $\mathbf{r}'$ across the beam propagation direction and $\xi_{x,y} = L_1/(k\sigma_{x,y})$ are usually defined as the two transverse coherence lengths. For a high brilliance 3$^{rd}$ generation synchrotron source as PETRA III [71] with the rms-source size $\sigma_x \approx 36$ µm, $\sigma_y \approx 6$ µm corresponding to a low-β operation of the storage ring at the distance $L_1 = 60$ m from the source and x-ray energy $E_\gamma \approx 12$ keV we obtain for the coherence length in both directions $\xi_x \approx 30$ µm and $\xi_y \approx 200$ µm [70].

We will further assume that the time autocorrelation function $F(\tau)$ in (41) has a pure exponential form

$$F(\tau) = F_0 \exp(-\tau/\tau_\parallel) \qquad (45)$$

which would be the exact result for a Lorentzian power spectral density of the source [66]. The characteristic time $\tau_\parallel$ of the decay of the time autocorrelation function defines the longitudinal coherence length $\xi_\parallel = c\tau_\parallel$. It can easily be shown [64-66] that the coherence length $\xi_\parallel$ is determined by the bandwidth $(\Delta\lambda/\lambda)$ of the incoming radiation and for exponential autocorrelation function $F(\tau)$ it is equal to $\xi_\parallel = (2/\pi)(\lambda^2/\Delta\lambda)$. For a Si(111) double-crystal monochromator with a bandwidth $\Delta\lambda/\lambda \approx 3\times 10^{-4}$ and the wavelength $\lambda \approx 1.5$ Å (which corresponds to $E_\gamma \approx 8$ keV) we get for the longitudinal coherence length $\xi_\parallel \approx 0.32$ µm.

In the far-field limit the time autocorrelation function $F(\Delta\tau)$ is given by

$$F(\Delta\tau) = F(|\mathbf{r}_{\parallel} - \mathbf{r}_{\parallel}'|) = F_0 \exp(-|l_r - l_{r'}|/\xi_{\parallel}) = F_0 \exp(-|\mathbf{r}_{\parallel} - \mathbf{r}_{\parallel}'|/\xi_{\parallel}) \qquad (46)$$

where $\mathbf{r}_{\parallel}$ and $\mathbf{r}_{\parallel}'$ are the components of $\mathbf{r}$ and $\mathbf{r}'$ along the beam and we have neglected small perpendicular contribution.

Substituting expressions (40)-(46) into (39) we now obtain for the intensity

$$I(\mathbf{Q}) = \frac{|F(\mathbf{h})|^2}{v^2} \iint s(\mathbf{r})s(\mathbf{r}')\gamma_{in}(\mathbf{r}_{\perp} - \mathbf{r}_{\perp}')F(\mathbf{r}_{\parallel} - \mathbf{r}_{\parallel}')e^{-i\mathbf{Q}\cdot(\mathbf{r}-\mathbf{r}')}d^3r d^3r' \qquad (47)$$

where the complex degree of coherence $\gamma_{in}(\mathbf{r}_{\perp} - \mathbf{r}'_{\perp})$ is defined by (44) and the autocorrelation function $F(\mathbf{r}_{\parallel} - \mathbf{r}'_{\parallel})$ by (46). This expression can further be simplified by changing the variables

$$I(\mathbf{Q}) = \frac{|F(\mathbf{h})|^2}{v^2} \int \varphi_{11}(\mathbf{r})\gamma_{in}(\mathbf{r}_{\perp})F(\mathbf{r}_{\parallel})e^{-i\mathbf{Q}\cdot\mathbf{r}}d^3r \qquad (48)$$

where $\varphi_{11}(\mathbf{r}) = \int s(\mathbf{r}')s(\mathbf{r}'+\mathbf{r})d^3r'$ is the autocorrelation function of the shape function $s(\mathbf{r})$.

In the coherent limit, the transverse and longitudinal coherence lengths $\xi_{\perp}$, $\xi_{\parallel}$ become infinite which as a consequence means $\gamma_{in}(\mathbf{r}_{\perp}) \to 1$ for the limit for the complex degree of coherence and $F(\mathbf{r}_{\parallel}) \to 1$ for the autocorrelation function. In this case we get for the intensity of the coherently scattered radiation

$$I_{coh}(\mathbf{Q}) = \frac{|F(\mathbf{h})|^2}{v^2} \int \varphi_{11}(\mathbf{r})e^{-i\mathbf{Q}\cdot\mathbf{r}}d^3r = |A(\mathbf{Q})|^2 \qquad (49)$$

where $A(\mathbf{Q}) = (F(\mathbf{h})/v)\int s(\mathbf{r})\exp(-i\mathbf{Q}\cdot\mathbf{r})d^3r$ is the kinematically scattered amplitude from the crystal with shape function $s(\mathbf{r})$. This result completely coincides (assuming zero strain field $\mathbf{u}(\mathbf{r})=0$) with the coherent limit of equations (30) and (31) discussed above.

Applying the convolution theorem, the intensity $I(\mathbf{Q})$ in equation (48) can be written in the form of convolution of two functions

$$I(\mathbf{Q}) = \frac{1}{(2\pi)^3}\int I_{coh}(\mathbf{Q}')\tilde{\Gamma}(\mathbf{Q}-\mathbf{Q}')d^3Q' = I_{coh}(\mathbf{Q}) \otimes \tilde{\Gamma}(\mathbf{Q}) \qquad (50)$$

where $I_{coh}(\mathbf{Q})$ is intensity of coherently scattered radiation (49) and $\tilde{\Gamma}(\mathbf{Q})$ is the Fourier transform

$$\tilde{\Gamma}(\mathbf{Q}) = \int \gamma_{in}(\mathbf{r}_\perp) F(|\mathbf{r}_\|)|)e^{-i\mathbf{Q}\cdot\mathbf{r}}d^3r \qquad (51)$$

Now we will consider orthogonal coordinates with the $z$ axis along the diffracted beam propagation direction and the $x,y$ axes perpendicular to this direction as already used in the previous sections. In this coordinate system for the exact Bragg position $\mathbf{q}=\mathbf{h}$ and $\mathbf{Q}=\mathbf{q}_v$ we can write intensity distribution (48) in the detector plane as

$$I(\mathbf{q}_v) = \frac{|F(\mathbf{h})|^2}{v^2}\int \varphi_{11}^z(\mathbf{x})\gamma_{in}(\mathbf{x})e^{-i\mathbf{q}_v\cdot\mathbf{x}}d^2x \qquad (52)$$

where $\mathbf{x}$ is a 2D vector $\mathbf{x}=(x,y)$ and $\varphi_{11}^z(\mathbf{x}) = \int \varphi_{11}(\mathbf{r})\exp(|-z|/\xi_\|)dz$. It is interesting to note, that the intensity distribution (52) can also be calculated as a convolution of the functions $\varphi_{11}^z(\mathbf{q})$ and $\gamma_{in}(\mathbf{q})$. For the case of large longitudinal length $\xi_\| \gg D$ the function $\varphi_{11}^z(\mathbf{q})$ gives just the projection of the 3D autocorrelation function to the plane $\mathbf{x}$. Smaller values of $\xi_\| \leq D$ reduce the real space volume of the scattering object along the propagating beam that contribute coherently in the diffraction pattern.

In Fig. 9 we present calculations of 2D diffraction patterns obtained from Eq. (52) for the crystal shape shown in the left inset of Fig. 5 with different values of transverse coherence lengths $\xi_x$ and $\xi_y$. For simplicity it was assumed here that the longitudinal coherence length is big enough $\xi_\parallel \gg D$ and it was not considered in the simulations. From this figure it is obvious, that decreasing the values of parameters $\xi_x$ and $\xi_y$ leads to a decrease in the contrast of the diffraction pattern. In the right column in Fig. 9 a reconstruction of the crystal projected electron density from the corresponding diffraction patterns is presented (see for details [29,30]). Not surprisingly, the quality of the reconstructions becomes poor with small degree of coherence.

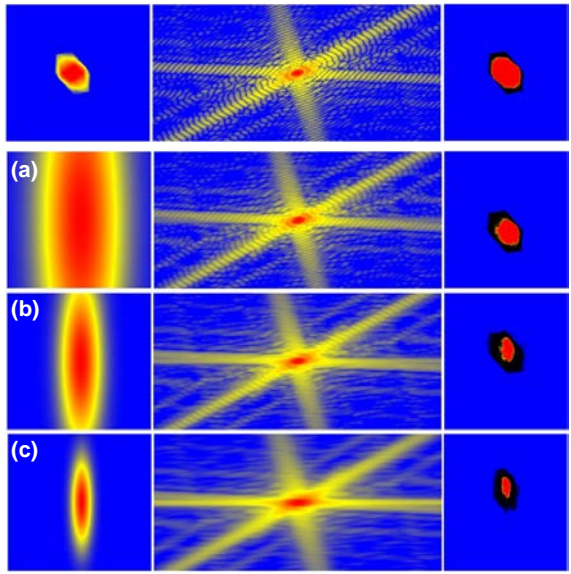

**Fig. 9** Complex degree of coherence $\gamma_{in}(x, y)$ (left column) used for calculations of diffraction intensity patterns (central column). Reconstructed real space images are shown in right column. For comparison on the top row shown the case of coherent illumination with the coherence length $\xi_x = \xi_y = \infty$. The values of the coherence lengths from top down: (a) $\xi_x = 91$ pixels, $\xi_y = 367$ pixels, (b) $\xi_x = 45$ pixels, $\xi_y = 183$ pixels, (c) $\xi_x = 22$ pixels, $\xi_y = 91$ pixels. Adapted from Ref. [29].

## 3. EXPERIMENTAL EXAMPLES

In this section we give several experimental examples of applications with coherent X-ray diffractive imaging to different systems.

### *3.1 Coherent X-ray imaging of defects in colloidal crystals*

First, we present results of CXDI applied to reveal the structure of a regular part and also a part containing a defect of a colloidal 2D crystal (see for details [53]).

Self-organized colloidal crystals are an attractive material for modern technological devices. They can be used as the basis for novel functional materials such as photonic crystals, which may find applications in future solar cells, LEDs, lasers or even as the basis for circuits in optical computing and communication. For these applications the crystal quality is crucial and monitoring the defect structure of real colloidal crystals is essential [72].

The experiment [53] was performed at the microoptics test bench at the ID06 beamline of the European Synchrotron Radiation Facility (ESRF) using an incident X-ray energy of 14 keV. The geometry of the experiment allows for rotation of the sample around the vertical axis which is perpendicular to the incident X-ray direction (see sketch Fig. 10). A 6.9 µm pinhole was positioned at a close distance in front of the colloidal crystal. The pinhole selects a highly coherent part of the beam and produces a finite illumination area. The initial orientation of the sample (with azimuthal angle $\varphi = 0°$) corresponds to the direction of the incident X-rays along the [111] direction of the fcc colloidal crystal and was perpendicular to the surface normal of the sample. Rotating the sample around the $x$-axis allows the measurement of different sets of diffraction planes. Particularly important was the direction of the incident X-rays along the [110] direction of the colloidal sample lattice at $\varphi = 35°$, when the set of (111) planes was aligned along the incident beam. The diffraction data were recorded using a CCD with 4005×2671 pixels with a resolution of $\Delta q = 0.16$ µm$^{-1}$ per pixel. In experiment a thin film of a colloidal crystalline sample on a glass substrate was used. It was grown by the convective assembly technique using polystyrene microspheres (diameter 425 nm, relative standard deviation 5%. The grown crystalline films have a face centered cubic (fcc) structure and were typically 20 to 30 layers thick.

Positioning the sample just after the aperture yields strong fringes typical of an Airy pattern [64] from a circular aperture, centered at $q=0$ (Fig. 11a,d). In addition, due to the long range order in the colloidal crystal, several orders of Bragg peaks are easily visible in the diffraction patterns. The strongest are the hexagonal set of 220 Bragg peaks typical for scattering from an fcc structure. Each of these Bragg peaks contains a few orders of diffraction fringes similar to those at $q=0$, due to the finite aperture in front of the sample. In addition to the allowed 220 Bragg peaks, we also observed much weaker forbidden peaks ($(2/3, 2/3, 4/3)$ in our case). Their appearance is an indication of defects in the crystal.

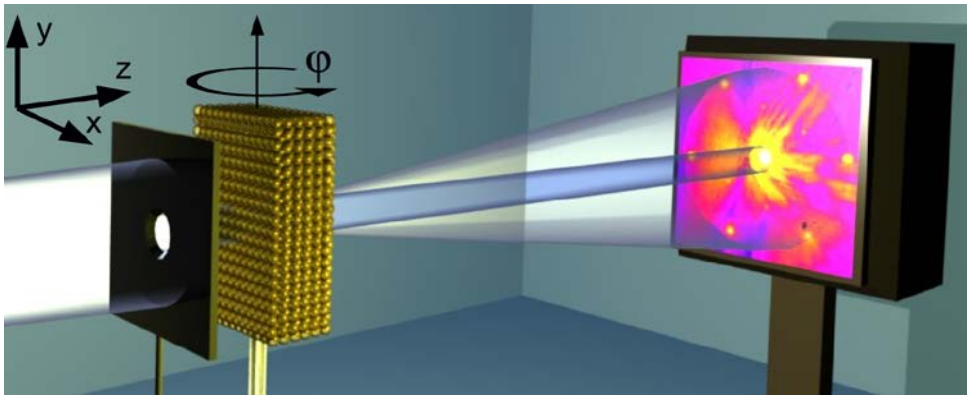

**Fig. 10** Schematic of the experiment with colloidal crystals. Adapted from Ref. [53].

The measured diffraction data were inverted by applying the guided hybrid input-output GHIO algorithm [17]. To improve the quality of the reconstruction a scaled diffraction pattern of the pinhole was subtracted from the diffraction pattern of the sample Fig. 11a. After this procedure the 220 Bragg peaks, and especially the fringes around them, are clearly visible against the background Fig. 11b. Negative values, shown in black in the difference diffraction pattern in Fig. 11b, were left to evolve freely in the reconstruction procedure. To stabilize the reconstruction process the central region (with $q < 5.44 \ \mu m^{-1}$) of the reconstructed diffraction pattern was kept fixed after 20 initial iterations.

A real space image of a colloidal sample obtained as a result of reconstruction of the diffraction pattern shown in Fig. 11b is presented in Fig. 11c. This image represents a *projection* of the 'atomic' structure of the colloidal crystal along the [111] direction. The hexagonal structure is clear across the whole illuminated region, with only slightly lower intensity values of the image around the edges of the pinhole aperture. As a consequence of the image being a projection of 3D arrangement of colloidal particles, the periodicity does not correspond to the colloidal interparticle distance $d$ in a *single* crystalline layer. Due to ABC ordering in fcc crystals a reduced periodicity of $d/\sqrt{3}$ is measured in this geometry.

The major differences from the results of previous work with CXDI on crystalline samples [44] are clearly demonstrated in Fig. 11c. Instead of a continuous shape and strain field reconstructed from the measurements of diffraction patterns around a single Bragg peak, the hexagonal structure shown in Fig. 11c gives the projected *positions* of the colloidal particles as it was discussed in details in section 2.1

An estimate of the resolution by performing line scans through the reconstructed image in Fig. 11c and by measuring the widths of the peaks by Gaussian fit gives a full width half maximum (FWHM) of 95 nm at the particle positions in Fig. 11c.

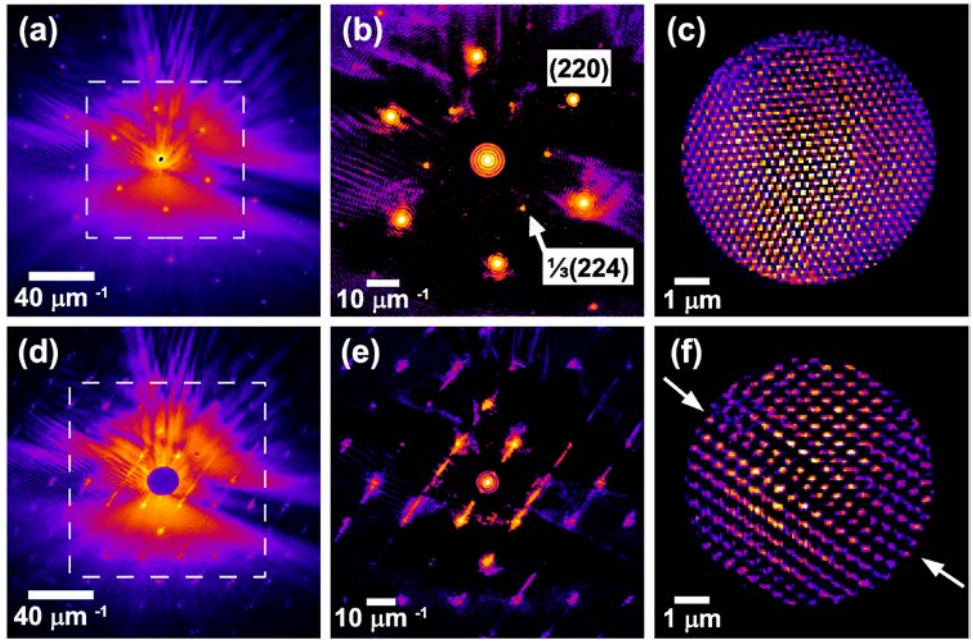

**Fig. 11** (a, d) Measured diffraction patterns from the pinhole and the sample. Marked regions in (a, d) correspond to the area used for the reconstruction. (b, e) Difference diffraction patterns obtained as a result of the subtraction of the scaled diffraction pattern of the pinhole from the measured diffraction patterns of the sample. Diffraction patterns are shown on a logarithmic scale. (c, f) Reconstruction of the colloid sample from the diffraction patterns. The arrows in (f) point to the defect in the crystal. First row (a,b,c) measurement at the azimuthal angle $\varphi = 0$ and the second row (d,e,f) at the angle $\varphi = 35°$. Adapted from Ref. [53].

The diffraction patterns measured at an angle of $\varphi = 35°$ (see Fig. 11d) were especially intriguing. They show strong streaks of varying intensity originating at the Bragg peaks with an angle of $55°$ to the horizontal direction (Fig. 11d,e). It is well known from previous studies [73] of similar colloidal systems that such streaks in reciprocal space are induced by stacking faults in the fcc structure in the (111) planes.

These diffraction patterns were reconstructed using the same procedure as described earlier and the result of this reconstruction is presented in Fig. 11f. The 'atomicity' of the colloidal crystal sample is again present in the reconstruction. In addition, a stacking fault appears (indicated by arrows in Fig. 11f) as a break in the 'correct' ABC ordering [33]. One can see a stacking fault, which consists of two hcp planes, and two fcc domains with the same stacking direction. The effect of the stacking fault is a translation of the two fcc crystals relative to each other. This 'sliding' can be seen in Fig. 11f as a 'break' of the lines of bright spots at the defect. It was recently suggested that, over a large (submillimeter) sample area, these double stacking defects consisting of two hcp planes is a common imperfection in convectively assembled colloidal crystals [74].

In this part we demonstrated that the simple and nondestructive mechanism of coherent X-ray diffractive imaging opens a unique route to determine the structure of mesoscopic materials such as colloidal crystals. CXDI has the potential to provide detailed information about the local defect structure in colloidal crystals. This is especially important for imaging photonic materials when refraction index matching is not possible or the sizes of colloidal particles are too small for conventional optical microscopy. To extend this method to larger fields of view scanning methods such as ptychography [18-20] can be used, while tomographic methods such as coherent X-ray tomography [75] have the potential to visualize the atomic structure of the defect core in 3D (see, for example, recent publication [54]).

## 3.2 Coherent diffraction tomography of nanoislands from grazing incidence small-angle X-ray scattering

In our second example we show how tomographic methods can be combined with CXDI to provide 3D images of nanocrystaline materials (see for details [75]).

Tomography and especially X-ray tomography has become one of the most important tools for investigating 3D structures in condensed matter [76]. When projected absorption contrast or phase contrast measurements are carried out in conventional X-ray transmission tomography the achievable resolution is limited by the spatial resolution of the area detector that can be about one micrometer. CXDI represents a possible solution to this dilemma. As no lenses are required in this imaging technique and the resolution is given by the scattered signal in principle the resolution limits of conventional X-ray transmission tomography can be overcome.

To obtain a 3D image of a non-crystallographic object in the forward scattering geometry by the CXDI technique the sample is usually mounted on a $Si_3N_4$ membrane (see for e.g. [77]) and then rotated with fixed azimuthal angular steps (see inset (a) in Fig. 12). Unfortunately, in this approach not all angles for a full 3D scan are accessible due to the positioning of the object on a membrane. Obviously, the inaccessible part of the reciprocal space is not available for the tomographic reconstruction, resulting in a certain loss of features which are actually present in the original object. Instead, as it was first proposed in [48,75], a sample can be positioned on a flat thick substrate and tomographic scans can be performed by collecting successive coherent scattering diffraction patterns at different azimuthal positions of a sample in a grazing-incidence small-angle X-ray scattering (GISAXS) geometry [48] (see Fig. 12). With this approach there are no limitations on the angle of rotation. Consequently large areas in reciprocal space can be measured with sufficient resolution and without missing wedge. The feasibility of this approach was tested and proven by a number of simulations [78]. Below experimental realization of this coherent diffraction tomographic technique is reported [75].

As a model samples SiGe islands of 200 nm average base size grown by liquid phase epitaxy were used. All islands were coherently grown on a (001) Si surface and exhibit a truncated pyramidal shape with a square base (see inset (b) in Fig. 12). In addition they exhibit a narrow size distribution (~10% full width at half maximum (FWHM)) and the same crystallographic orientation on the Si surface (Fig. 12).

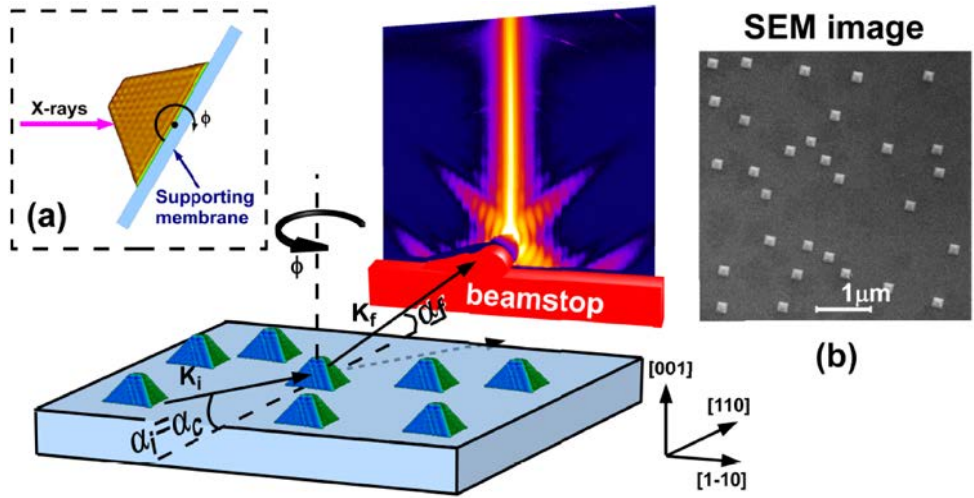

**Fig. 12** Schematic diagram of the GISAXS scattering geometry on a group of nano-islands in the form of a truncated pyramid with a square base. The incident wave vector $\mathbf{k}_i$ at grazing incidence angle $\alpha_i = \alpha_c$ is shown and the scattered wave vector $\mathbf{k}_f$ at angles $\alpha_f \geq \alpha_c$. The sample is rotated around the surface normal (azimuth angles). Inset (a): Schematic diagram of a conventional CXDI tomography when the sample is positioned on a supporting membrane. Inset (b): A scanning electron microscopy image of the nano-islands. Adapted from Ref. [75].

Experiments [48,75,79] were performed at the ID01 beamline of the European Synchrotron Radiation Facility (ESRF) in Grenoble. The incidence angle was taken equal to the critical angle for total external reflection of the Si substrate which corresponds to $\alpha_i = \alpha_c = 0.224°$ for the chosen X-ray energy of 8 keV. This particular angle was used because at these conditions the scattering may be considered as predominantly kinematical [80]. The coherently scattered signal was measured up to $q_\parallel = \pm 0.56 \text{ nm}^{-1}$ in reciprocal space in the transverse direction. However, due to a certain noise level only a limited part of reciprocal space up to $q_\parallel = \pm 0.36 \text{ nm}^{-1}$ was considered for the reconstruction, which provides a 17.4 nm resolution in real space. An azimuth scan was performed from 5° to 50° with an angular increment of 1°. Due to the four fold and mirror symmetry of {111} facetted islands such scans cover the whole reciprocal space. During the azimuthal scan the incidence angle was kept constant at the critical angle $\alpha_c$.

The diffraction patterns at each azimuthal angle position represent slices through the 3D reciprocal space on a pseudopolar grid (see Fig. 13), and can be combined to produce a 3D intensity distribution in reciprocal space. An iso-intensity surface of this 3D representation of the scattered intensity is shown in Fig. 13c. Strong crystal truncation rods (CTR) along the <111> directions and much weaker crystal truncation planes (CTP) connecting the CTRs can be observed [48]. For comparison we performed calculations of GISAXS diffraction patterns in the framework of the distorted-wave Born-approximation (DWBA) theory [81,82] for similar pyramids. The corresponding 3D representation of the scattered intensity is shown in Fig. 13a. Strong interference fringes due to the coherent scattering of the X-ray beam from a small pyramid-shaped object can be observed in this figure. These interference fringes are smeared out in our experimental data (Fig. 13c) partly due to a finite size distribution of the islands and partly due to a lack of sufficient counting statistics. In order to model the experimental results we added a Gaussian mask to the calculated set of data. The corresponding 3D representation of the scattered intensity is shown in Fig. 13b.

Results of the island shape reconstruction from the experimental

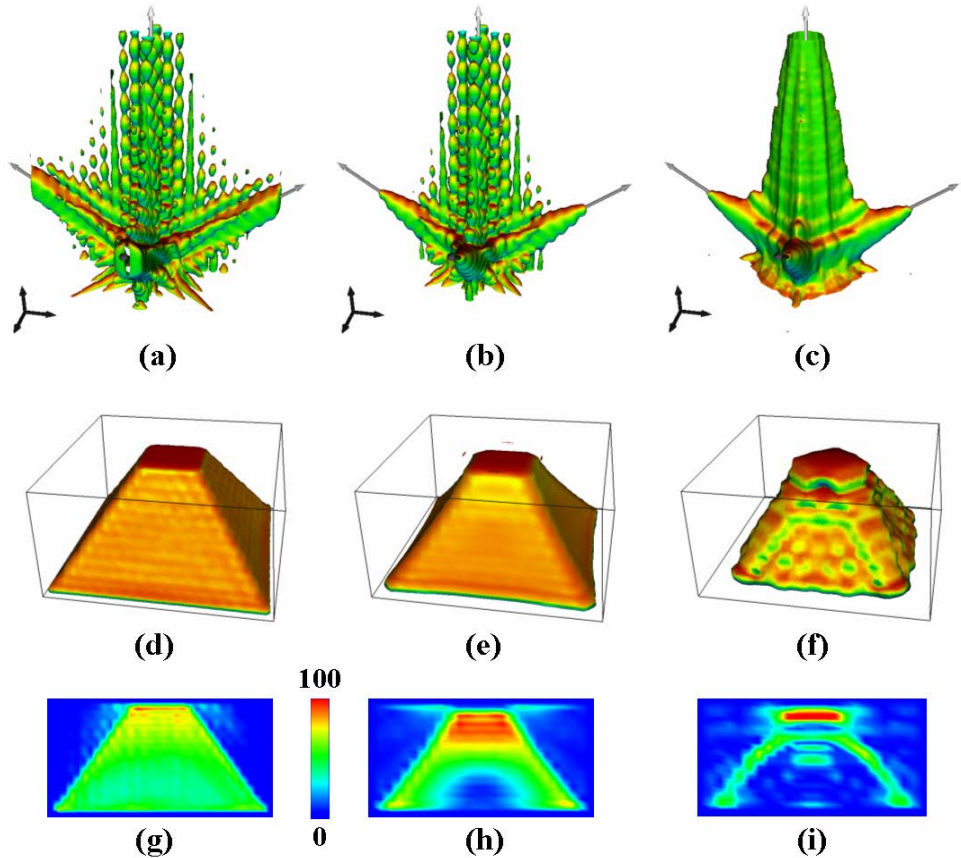

**Fig. 13** Left column [(a), (d), (g)]: simulations in the framework of the DWBA theory. Middle column [(b), (e), (h)]: simulations in the framework of the DWBA theory with an additional Gaussian mask (see text for details). Right column [(c), (f), (i)]: experiment. [(a), (b),(c)]: 3D plot of an iso-intensity surface in the reciprocal space. RGB colors correspond to the z-projection of the iso-surface normal. Grey arrows indicate directions along the crystallographic planes (001) top and {111} on the side. Black arrows indicate $q_x, q_y, q_z$ directions in reciprocal space. The length of each black arrow corresponds to 0.1 $nm^{-1}$. [(d), (e), (f)]: Reconstructed shape of the islands. The transparent box indicates the size of the support. [(g), (h), (i)]: Electron density of the islands obtained as a vertical section through the center of each island. Adapted from Ref. [75].

GISAXS diffraction patterns are presented in Fig. 13f. For a comparison results of the reconstruction of the island shape from simulated data are also shown in Fig. 13d,e. The electron densities of the islands obtained as a vertical section through the center of the islands are presented in Fig. 13g-i. From these results it is seen that the shape of the islands is reconstructed correctly for the experimental data set (Fig. 13f). However, for the electron density inside the island we observe artifacts in the form of low density regions in the bottom of the islands (Fig. 13i). Reconstructions performed with the simulated data sets show that the scattering data obtained in the DWBA conditions correctly reproduce the shape (Fig. 13d) and electron density (Fig. 13g) of an island. However, when the modified theoretical data set with the Gaussian mask is used for reconstruction, artifacts similar to those from the experimental data set appear. These results suggest that the artifacts can be removed by an increased incidence flux (e.g. by using focusing optics [22]) and with the use of a new generation of detectors with extremely high dynamic range.

Here we have demonstrated how this approach of coherent diffraction GISAXS can be used to obtain the 3D electron density of nanometer sized islands. This was achieved by performing tomographic azimuth scans in a GISAXS geometry on many identical islands and subsequent phase retrieval which yields the tomographic information, such as the shape and the electron density. It is important to note that this approach does not depend on the crystalline structure of such an island and may be applied to any material system.

### *3.3 Coherent-pulse 2D crystallography at free electron lasers*

As our third example we show how ultra-bright coherent pulses of a new free-electron laser (FEL) sources can be applied to determine the structure of two-dimensional (2D) crystallographic objects (see for details Ref. [83], and also for reviews of CXDI experiments at FEL source [5,6,9]).

Crystallization and radiation damage is presently a bottleneck in protein structure determination. As it was first proposed in Ref. [83] two-dimensional (2D) finite crystals and ultra-short FEL pulses can be effectively used to reveal the structure of single molecules. This can be especially important for membrane proteins that in general do not form 3D crystals, but easily form 2D crystalline structures. In this paper single pulse train coherent diffractive imaging was demonstrated for a finite 2D crystalline sample, and it was concluded that this alternative approach to single molecule imaging is a significant step towards revealing the structure of proteins with sub-nanometer resolution at the newly built XFEL sources.

Revealing the structure of protein molecules is mandatory for understanding the structure of larger biological complexes. The major progress in uncovering the structure of proteins in past decades was due to the development of phasing methods [84] allowing the determination of the structure of complex molecules that crystallize. One new approach to overcome these difficulties is based on the use of ultra-short pulses of X-ray free-electron lasers (XFEL) [85-87]. This elegant idea is based on measuring a sufficiently sampled diffraction pattern from a single molecule illuminated by an FEL pulse [88,89]. However, in spite of the extreme intensity of the FEL pulses, a diffraction pattern from only one molecule will not be sufficient to obtain a high resolution diffraction pattern. Many reproducible copies will need to be measured to get a sufficient signal to noise ratio for each projection necessary for three-dimensional (3D) imaging at sub-nanometer spatial resolution.

Free-electron lasers are especially well suited for such coherent 2D crystallography. They provide femtosecond coherent pulses [70,90,91] with extremely high power. Only the combination of all of these unique properties will allow the realization of 2D crystallographic X-ray imaging on biological systems. Brilliant, ultra-short pulses could overcome the radiation damage problem [88,92] which is a severe limitation of conventional crystallography at 3rd generation synchrotron sources [93]. Higher luminosity and hence improved statistics for such experiments can be obtained by the use of pulse trains that can be provided by FLASH (Free-electron LASer in Hamburg) [94].

A finite 2D crystallography was demonstrated by using a micro-structured crystal array that was prepared on a 100 nm thick silicon nitride membrane substrate coated with 600 nm of gold, and 200 nm of palladium. The finite crystal sample was manufactured by milling holes in the film in a regular array pattern using a Focused Ion Beam (FIB). The 'unit cell' of our crystal consists of a large hole of 500 nm diameter (representing a 'heavy atom' in conventional crystallography) and a smaller hole of 200 nm diameter (representing a 'light atom'). The whole structure was composed of five unit cells in each direction, making the total structure size about 10 µm x 10 µm.

The diffraction data were measured at FLASH on the PG2 monochromator beamline [95] with a fundamental wavelength of 7.97 nm. An exposure time of 0.2 s was used to collect a series of single pulse train data from our sample. FLASH was operated in a regime producing 21 bunches of electrons per pulse train, with a pulse train repetition rate of 5 Hz. The bunches within each pulse train were spaced at 1 MHz. The average pulse energy was 15 µJ which is equivalent to $6 \times 10^{11}$ photons per pulse or $1.3 \times 10^{13}$ photons per train at the source. The coherent flux on the sample area was $1.5 \times 10^{10}$ photons per pulse train.

A typical data set is shown in Fig. 14a. The diffraction pattern as measured fills the whole detector, which corresponds to a minimum feature size of 220 nm (Fig. 14a). We note that all expected features of a finite, crystalline structure as they were discussed in the previous sections are observed in this diffraction pattern. The Bragg peaks due to the regular array are clearly seen, as are the oscillations between the Bragg peaks that are the result of the finite extent and coherent illumination of our sample. Also seen is the form factor from the individual elements – the large holes

– that can be observed as a radial intensity modulation across the pattern produced.

Due to the limited signal to noise ratio of the initial data set and the symmetry of the unit cell the initial reconstructions stagnated with two equivalent solutions superimposed. One is with the small dots appearing to the top right of the larger dots, and the other is with them appearing to the

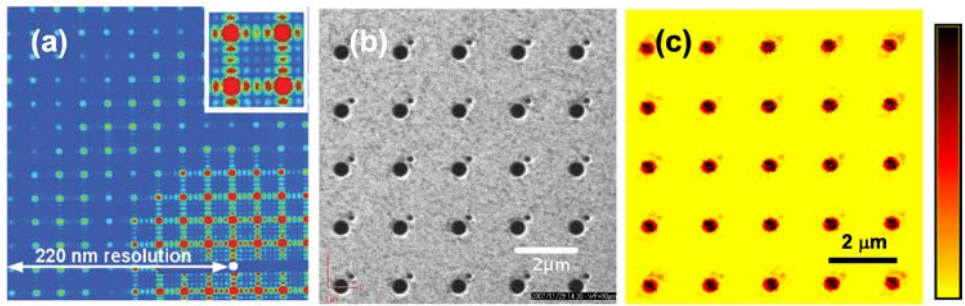

**Fig. 14** (a) Far-field diffraction data measured from a single train of 21 femtosecond pulses from the FEL. (Inset: Enlarged region of diffraction pattern). (b) SIM image of the finite, periodic structure. (c) The reconstructed image using the original data binned 5×5. Adapted from Ref. [83].

bottom left. To solve this problem the data were binned 5×5 and a more constrained support of 25 rectangular boxes each centered on the positions of the unit cell was used. This data set was used for reconstruction by applying the HIO [12] iterative phase retrieval algorithm. By increasing the signal-to-noise ratio and reducing the symmetry in real space it was possible to improve the reconstruction to resolve the smallest features in the sample (Fig. 14c). A scanning ion micrograph (SIM) image of the object under investigation is shown in Fig. 14,b for comparison. The resolution in real space in the reconstructed image was better than 240 nm. This compares favorably with the measured maximum momentum transfer corresponding to a 220 nm resolution.

Summarizing the XFEL experiment, it was demonstrated that single pulse train coherent diffractive imaging is possible for a finite 2D crystalline sample with the reconstructed image exhibiting resolution commensurate with the measured data. In this experiment the *crystalline* structure was essential in providing the necessary information to determine the structure of the unit cells. If only a *single* unit cell would have been used simulations suggest that a successful reconstruction would be impossible with the resolution presented in the example. This approach to use a pattern of single molecules is a significant step towards revealing the structure of proteins with sub-nanometer resolution at the newly built XFEL sources.

## 4. SUMMARY

Coherent X-ray diffractive imaging gives us a high resolution imaging tool to reveal the electron density and strain in nano-crystalline samples. Progress is still ongoing. We foresee that in future it will reach a resolution of approximately a few nanometers at synchrotron sources, and a few angstroms for the protein nano-crystals imaged with the ultrashort FEL pulses [96]. Non-reproducible objects may be imaged with a few nanometer resolution at XFELs [9,26,97]. There are several technological developments that are incremental for the future progress of the CXDI technique. The ultimate hard X-ray storage rings [98] with a few tens picometer emittance are under discussion. The advances in hard X-ray optics will allow reaching sub-ten nanometer focus sizes [99]. Also, hard X-ray FEL sources now become online. Finally, the mathematical tools for interpretation of data from coherent X-ray scattering experiments improve continuously. Thus, we can expect that CXDI will become eminently important for the investigation of the structure of materials on the nanoscale and will grow rapidly over the near future.

The authors acknowledge a careful reading of the manuscript by O. Seeck and B. Murphy.

# References

1. Vartanyants, I. A., Pitney, J. A., Libbert, J. L. and Robinson, I. K. (1997). Reconstruction of Surface Morphology from Coherent X-ray Reflectivity. *Phys. Rev. B*, **55**, pp. 13193-13202.
2. Miao, J., Charalambous, P., Kirz, J. and Sayre, D. (1999). Extending the methodology of X-ray crystallography to allow imaging of micrometre-sized non-crystalline specimens. *Nature*, **400**, pp. 342-344.
3. Robinson, I. K., Libbert, J. L., Vartanyants, I. A., Pitney, J. A., Smilgies, D. M., Abernathy, D. L., and Grübel, G. (1999). Coherent X-ray Diffraction Imaging of Silicon Oxide Growth. *Phys. Rev. B*, **60**, pp. 9965-9972.
4. Robinson, I. K., Vartanyants, I. A., Williams, G. J., Pfeifer, M. A. and Pitney, J. A. (2001). Reconstruction of the Shapes of Gold Nanocrystals using Coherent X-ray Diffraction. *Phys. Rev. Lett.*, **87**, pp. 195505.
5. Vartanyants, I. A., Robinson, I. K., McNulty, I., David, C., Wochner, P. and Tschentscher, Th. (2007). Coherent X-ray scattering and lensless imaging at the European XFEL Facility. *J. Synchrotron Rad.*, **14**, pp. 453-470.
6. Vartanyants, I. A., Mancuso, A. P., Singer, A., Yefanov, O. M. and Gulden, J. (2010). Coherence Measurements and Coherent Diffractive Imaging at FLASH. *J. Phys. B: At. Mol. Opt. Phys.*, **43**, pp. 194016/1-10.
7. Nugent, K.A. (2010). Coherent methods in the X-ray sciences. *Advances in Physics* **59**, 1-99.
8. Robinson, I. K. and Harder, R. (2009). Coherent X-ray diffraction imaging of strain at the nanoscale. *Nature Mater.*, **8**, pp. 291-298.
9. Mancuso, A. P., Yefanov, O. M. and Vartanyants, I. A. (2010). Coherent diffractive imaging of biological samples at synchrotron and free electron laser facilities. *J. Biotechnol.*, **149**, pp. 229-237.
10. Quiney, H. M. (2011). Coherent diffractive imaging using short wavelength light sources. *J. Mod. Opt.*, **57**, pp. 1109-1149.
11. Gerchberg, R. W. and Saxton, W. O. (1972). A Practical Algorithm for the Determination of Phase from Image and Diffraction Plane Pictures. *Optik*, **35**, pp. 237-246.
12. Fienup, J. R. (1982). Phase retrieval algorithms: a comparison. *Appl. Opt.*, **21**, pp. 2758-2769.
13. Elser, V. (2003). Phase retrieval by iterated projections. *J. Opt. Soc. Am. A*, **20**, pp. 40-55.
14. Marchesini, S. (2007). A unified evaluation of iterative projection algorithms for phase retrieval. *Rev. Sci. Instr.*, **78**, pp. 049901.
15. Bates, R. H. T. (1982). Fourier phase problems are uniquely solvable in more than one dimension. I: Underlying theory. *Optik*, **61**, pp. 247-262.
16. Shannon, C. E. (1949). Communication in the Presence of Noise. *Proc. IRE*, **37**, pp. 10-21.
17. Chen, C.-C., Miao, J., Wang, C. W. and Lee, T. K. (2007). Application of optimization technique to noncrystalline X-ray diffraction microscopy: Guided hybrid input-output method. *Phys. Rev. B*, **76**, pp. 064113.
18. Rodenburg, J. M., Hurst, A. C., Cullis, A. G., Dobson, B. R., Pfeiffer, F., Bunk, O., David, C., Jefimovs, K. and Johnson, I. (2007). Hard-X-Ray Lensless Imaging of Extended Objects. *Phys. Rev. Lett.*, **98**, pp. 034801.
19. Thibault, P., Dierolf, M., Menzel, A., Bunk, O., David, Ch. and Pfeiffer, F. (2008). High-Resolution Scanning X-ray Diffraction Microscopy. *Science*, **321**, pp. 379-382.
20. Schropp, A., Boye, P., Feldkamp, J. M., Hoppe, R., Patommel, J., Samberg, D., Stephan, S., Giewekemeyer, K., Wilke, R. N., Salditt, T., Gulden, J., Mancuso, A. P., Vartanyants, I. A., Weckert, E., Schöder, S., Burghammer, M., and Schroer C. G. (2010). Hard x-ray nanobeam characterization by coherent diffraction microscopy. *Appl. Phys. Lett.*, **96**, pp. 091102/1-3.

21. Godard, P., Carbone, G., Allain, M., Mastropietro, F., Chen, G., Capello, L., Diaz, A., Metzger, T. H., Stangl J. and Chamard, V. (2011). Three-dimensional high-resolution quantitative microscopy of extended crystals. *Nature Comm.*, **2**, pp. 568.
22. Takahashi, Y., Zettsu, N., Nishino, Y., Tsutsumi, R., Matsubara, E., Ishikawa, T. and Yamauchi, K. (2010). Three-Dimensional Electron Density Mapping of Shape-Controlled Nanoparticle by Focused Hard X-ray Diffraction Microscopy. *Nano Lett.*, **10**, pp. 1922-1926.
23. Giewekemeyer, K., Pierre Thibault, P., Kalbfleisch, S., Beerlink, A., Kewish, C. M., Dierolf, M., Pfeiffer, F. and Salditt T. (2010). Quantitative biological imaging by ptychographic X-ray diffraction microscopy. *Proc. Natl. Acad. Sci. USA*, **107**, pp. 529-534.
24. Dierolf, M., Menzel, A., Thibault, P., Schneider, P., Kewish, C. M., Wepf, R., Bunk O. and Pfeiffer F. (2010). Ptychographic X-ray computed tomography at the nanoscale. *Nature*, **467**, pp. 436-439.
25. Shen, Q., Bazarov, I. and Thibault, P. (2004). Diffractive imaging of nonperiodic materials with future coherent X-ray sources. *J. Synchrotron Rad.*, **11**, pp. 432-438.
26. Bergh, M., Huldt, G., Timneanu, N., Maia, F. R. N. C. and Hajdu, J. (2008). Feasibility of imaging living cells at subnanometer resolutions by ultrafast X-ray diffraction. *Q. Rev. Biophys.*, **41**, pp. 181-204.
27. Huang, X., Miao, H., Steinbrener, J., Nelson, J., Shapiro, D., Stewart, A., Turner, J. and Jacobsen, Ch. (2009). Signal-to-noise and radiation exposure considerations in conventional and diffraction microscopy. *Optics Express*, **17**, pp. 13541-13553.
28. Thibault, P., Elser, V., Jacobsen, C., Shapiro D. and Sayre D. (2006). Reconstruction of a yeast cell from X-ray diffraction data. *Acta Cryst. A*, **62**, pp. 248-261.
29. Vartanyants, I. A. and Robinson. I. K. (2001). Partial coherence effects on the imaging of small crystals using coherent X-ray diffraction. *J. Phys.: Condens. Matter*, **13**, pp. 10593-10611.
30. Vartanyants, I. A. and Robinson, I. K. (2003). Origins of decoherence in coherent X-ray diffraction experiments. *Opt. Commun.*, **222**, pp. 29-50.
31. Williams, G. J., Quiney, H. M., Peele, A. G. and Nugent, K. A. (2007). Coherent diffractive imaging and partial coherence. *Phys. Rev. B*, **75**, pp. 104102.
32. Millane, R. P. (1990). Phase retrieval in crystallography and optics. *J. Opt. Soc. Am. A*, **7**, pp. 394-411.
33. Warren, B. E. (1990). *X-ray Diffraction*. (Dover Publ. Inc., New York).
34. Als-Nielsen, J. and McMorrow, D. (2011). *Elements of modern X-ray physics*. (John Wiley and Sons, New York).
35. Batterman, B. W., and Cole H. (1964). Dynamical Diffraction of X Rays by Perfect Crystals. *Rev. Mod. Phys.*, **36**, pp. 681-717.
36. Authier A. (2003). *Dynamical theory of X-ray diffraction*. (Oxford University Press, 2-nd edition, Oxford).
37. Shabalin, A., Yefanov, O. M., Nosik, V. L. and Vartanyants, I. A. Dynamical effects in coherent X-ray scattering on a finite crystal. (In preparation).
38. Harder, R., Pfeifer, M. A., Williams, G. J., Vartanyants, I. A. and Robinson, I. K. (2007). Orientation variation of surface strain. *Phys. Rev. B*, **76**, pp. 115425/1-4.
39. Ewald, P. P. (1940). X-ray diffraction by finite and imperfect crystal lattices. *Proc. Phys. Soc.*, **52**, pp. 167.
40. von Laue, M. (1936). Die äußere Form der Kristalle in ihrem Einfluß auf die Interferenzerscheinungen an Raumgittern. *Ann. d. Physik*, **26**, pp. 55.
41. Sayre, D. (1952). Some implications of a theorem due to Shannon. *Acta Cryst.*, **5**, pp. 843.
42. Williams, G. J., Pfeifer, M. A., Vartanyants, I. A. and Robinson, I. K. (2003). Three-Dimensional Imaging of Microstructure in Au Nanocrystals. *Phys. Rev. Lett.*, **90**, pp. 175501.
43. Millane, R. P. (1996). Multidimensional phase problems. *J. Opt. Soc. Am. A*, **13**, pp. 725.

44. Pfeifer, M. A., Williams, G. J., Vartanyants, I. A., Harder, R. and Robinson, I. K. (2006). Three-dimensional mapping of a deformation field inside a nanocrystal. *Nature,* **442**, pp, 63-66.
45. Robinson, I. K. (1986). Crystal truncation rods and surface roughness. *Phys. Rev. B*, **33**, pp. 3830-3836.
46. Afanas'ev, A. M., Aleksandrov, P. A., Imamov, R. M., Lomov, A. A. and Zavyalova, A. A. (1985). Three-Crystal Diffractometry in Grazing Bragg-Laue Geometry. *Acta Cryst. A*, **41**, pp. 227.
47. Pitney, J. A., Robinson, I. K., Vartanyants, I. A., Appelton, R. and Flynn, C. P. (2000).Streaked speckle in $Cu_3Au$ coherent diffraction. *Phys. Rev. B* **62**, 13084-13088.
48. Vartanyants, I. A., Zozulya, A. V., Mundboth, K., Yefanov, O. M., Richard, M.-I., Wintersberger, E., Stangl, J., Diaz, A., Mocuta, C., Metzger, T. H., Bauer, G., Boeck, T. and Schmidbauer, M. (2008). Crystal truncation planes revealed by three-dimensional reconstruction of reciprocal space. *Phys. Rev. B* **77**, 115317.
49. Eisebitt, S., Luning, J., Schlotter, W. F., Lorgen, M., Hellwig, O., Eberhardt, W. and Stohr, J. (2004). Lensless imaging of magnetic nanostructures by X-ray spectro-holography. *Nature* **432**, 885-888.
50. Stadler, L.-M., Gutt, Ch., Autenrieth, T., Leupold, O., Rehbein, S., Chushkin, Y. and Grübel, G. (2008). Hard X-Ray Holographic Diffraction Imaging. *Phys. Rev. Lett.* **100**, 245503.
51. Chamard, V., Stangl, J., Carbone, G., Diaz, A., Chen, G., Alfonso, C., Mocuta, C. and Metzger, T. H. (2010), Three-Dimensional X-Ray Fourier Transform Holography: The Bragg Case. *Phys. Rev. Lett.* **104**, 165501.
52. Gulden, J., Yefanov, O. M., Weckert, E., and Vartanyants, I.A. (2011). Imaging of Nanocrystals with Atomic Resolution Using High-Energy Coherent X-rays. *The 10th International Conference on X-ray Microscopy, AIP Conf. Proc.* **1365**, 42-45.
53. Gulden, J., Yefanov, O. M., Mancuso, A. P., Abramova, V. V., Hilhorst, J., Byelov, D., Snigireva, I., Snigirev, A., Petukhov, A. V. and Vartanyants, I. A. (2010). Coherent X-ray imaging of defects in colloidal crystals. *Phys. Rev. B* **81**, 224105.
54. Gulden, J., Yefanov, O. M., Mancuso, A. P., Dronyak, R., Singer, A., Bernátová, V., Burkhardt, A., Polozhentsev, O., Soldatov, A., Sprung, M. and Vartanyants I. A., (2012). Three-dimensional structure of a single colloidal crystal grain studied by coherent x-ray diffraction. *Optics Express*, **20**, pp. 4039-4049.
55. Zuo, J.-M., Vartanyants, I. A., Gao, M., Zhang, R. and Nagahara, L. A. (2003). Atomic resolution imaging of a carbon nanotube from diffraction intensities. *Science*, **300**, pp. 1419-1421.
56. Huang, W. J., Zuo, J. M., Jiang, B., Kwon, K. W. and Shim, M. (2009). Sub-angstrom-resolution diffractive imaging of single nanocrystals. *Nat. Phys.*, **5**, pp. 129-133.
57. Dronyak, R., Liang, K. S., Stetsko Y. P., Lee, T.-K., Feng C.-K., Tsai, J.-S. and Chen, F.-R. (2009). Electron diffractive imaging of nano-objects using a guided method with a dynamic support. *Appl. Phys. Lett.*, **95**, pp. 111908.
58. Krivoglaz, M. A. (1969). Theory of X-ray and Thermal-Neutron Scattering by Real Crystals. (Plenum Press, New York).
59. Robinson, I. K and Vartanyants, I. A. (2001). Use of Coherent X-ray Diffraction to Map Strain Fields in Nanocrystals. *Appl. Surf. Sci.*, **182**, pp. 186-191.
60. Vartanyants, I. A., Ern, C., Donner, W., Dosch, H., and Caliebe W. (2000). Strain profiles in epitaxial films from x-ray Bragg diffraction phases. *Appl. Phys. Lett.*, **77**, pp. 3929-3931.
61. Minkevich, A. A., Gailhanou, M., Micha, J.-S., Charlet, B., Chamard, V. and Thomas, O. (2007). Inversion of the diffraction pattern from an inhomogeneously strained crystal using an iterative algorithm, *Phys. Rev. B*, **76**, pp. 104106/1-5.

62. Newton, M. C., Leake, S. J., Harder R. and Robinson, I. K. (2010). Three-dimensional imaging of strain in a single ZnO nanorod. *Nature Mater.*, **9**, pp. 120-124.
63. Watari, M., McKendry, R. A., Vögtli, M., Aeppli, G., Soh, Y.-A., Shi, X., Xiong, G., Huang, X., Harder, R. and Robinson, I. K. (2011). Differential stress induced by thiol adsorption on facetted nanocrystals, *Nature Mater.*, **10**, pp. 862-866.
64. Born, M. and Wolf, E. (2000). *Principles of Optics*. (Cambridge University Press, Cambridge).
65. Mandel, L. and Wolf, E. (1995). *Optical Coherence and Quantum Optics*. (Cambridge University Press, Cambridge).
66. Goodman, J. W. (1985). *Statistical Optics*. (Wiley, New York).
67. Szöke, A. (2001). Diffraction of partially coherent X-rays and the crystallographic phase problem. *Acta Cryst.*, A**57**, pp. 586-603.
68. Sinha, S. K., Tolan, M. and Gibaud, A. (1998). The Effects of Partial Coherence on the Scattering of X-Rays by Matter. *Phys. Rev. B*, **57**, pp. 2740-2758.
69. Gutt, C., Ghaderi, T., Tolan, M., Sinha, S. K., and Grübel G. (2008). Effects of partial coherence on correlation functions measured by x-ray photon correlation spectroscopy. *Phys. Rev. B*, **77**, pp. 094133/1–10.
70. Vartanyants, I. A. and Singer, A. (2010). Coherence Properties of Hard X-Ray Synchrotron Sources and X-Ray Free-Electron Lasers. *New J. Phys.* **12**, 035004.
71. Balewski, K., Brefeld, W., Decking, W., Franz, H., Röhlsberger, R. and Weckert, E. (2004). *PETRA III: A Low Emittance Synchrotron Radiation Source. Technical Design Report.* (Hamburg, Germany: DESY).
72. Lodahl, P., van Driel, A. F., Nikolaev, I. S., Irman, A., Overgaag, K., Vanmaekelbergh, D. and Vos, W. L. (2004). Controlling the dynamics of spontaneous emission from quantum dots by photonic crystals. *Nature*, **430**, pp. 654.
73. Hilhorst, J., Abramova, V. V., Sinitskii, A., Sapoletova, N. A., Napolskii, K. S., Eliseev, A. A., Byelov, D. V., Grigoryeva, N. A., Vasilieva, A. V., Bouwman, W. G., Kvashnina, K., Snigirev, A., Grigoriev, S. V. and Petukhov, A. V. (2009). Double Stacking Faults in Convectively Assembled Crystals of Colloidal Spheres. *Langmuir*, **25**, pp. 10408.
74. Napolskii, K. S., Sapoletova, K. S., Gorozhankin, D. F., Eliseev, A. A., Chernyshov, D. Y., Byelov, D. V., Grigoryeva, N. A., Mistonov, A. A., Bouwman, W. G., Kvashnina, K. O., Lukashin, A. V., Snigirev, A. A., Vassilieva, A. V., Grigoriev, S. V. and Petukhov, A. V. (2010). Fabrication of artificial opals by electric-field-assisted vertical deposition. *Langmuir*, **26**, pp. 2346.
75. Yefanov, O. M., Zozulya, A. V., Vartanyants, I. A., Stangl, J., Mocuta, C., Metzger, T. H., Bauer, G., Boeck, T. and Schmidbauer, M. (2009). Coherent diffraction tomography of nanoislands from grazing-incidence small-angle X-ray scattering. *Appl. Phys. Lett.*, **94**, pp. 123104.
76. Banhart, J. (2008). Advanced Tomographic Methods in Materials Research and Engineering. (Oxford University Press, New York).
77. Miao, J., Chen, C.-C., Song, Ch., Nishino, Y., Kohmura, Y., Ishikawa, T., Ramunno-Johnson, D., Lee, T.-K. and Risbud, S. H. (2006). Three-Dimensional GaN-$Ga_2O_3$ Core Shell Structure Revealed by X-Ray Diffraction Microscopy. *Phys. Rev. Lett.*, **97**, pp. 215503.
78. Yefanov, O. M. and Vartanyants, I. A. (2009). Three dimensional reconstruction of nanoislands from grazing-incidence small-angle X-ray scattering. *Eur. Phys. J. Special Topics*, **167**, pp. 81-86.
79. Zozulya, A. V., Yefanov, O. M., Vartanyants, I. A., Mundboth, K., Mocuta, C., Metzger, T. H., Stangl, J., Bauer, G., Boeck, T. and Schmidbauer, M. (2008) Imaging of nano-islands in coherent grazing-incidence small-angle X-ray scattering experiments. *Phys. Rev. B*, **78**, pp. 121304.

80. Vartanyants, I., Grigoriev, D. and Zozulya, A. (2007). Coherent X-ray imaging of individual islands in GISAXS geometry. *Thin Solid Films*, **515**, pp. 5546.
81. Sinha, S. K., Sirota, E. B., Garoff, S. and Stanley, H. B. (1988). X-ray and neutron scattering from rough surfaces. *Phys. Rev. B*, **38**, pp. 2297.
82. Rauscher, M., Paniago, R., Metzger, H., Kovats, Z., Domke, J., Peisl, J., Pfannes, H.-D., Schulze, J. and Eisele, I. (1999). Grazing incidence small angle x-ray scattering from free-standing nanostructures. *J. Appl. Phys.*, **86**, pp. 6763.
83. Mancuso, A. P., Schropp, A., Reime, B., Stadler, L.-M., Singer, A., Gulden, J., Streit-Nierobisch, S., Gutt, C., Grübel, G., Feldhaus, J., Staier, F., Barth, R., Rosenhahn, A., Grunze, M., Nisius, T., Wilhein, T., Stickler, D., Stillrich, H., Frömter, R., Oepen, H.-P., Martins, M., Pfau, B., Günther, C. M., Könnecke, R., Eisebitt, S., Faatz, B. and Guerassimova, N. (2009). Coherent-Pulse 2D Crystallography Using a Free-Electron Laser X-Ray Source. *Phys. Rev. Lett.*, **102**, pp. 035502.
84. Woolfson, M. and Hai-fu, F. (1995). *Physical and non-physical methods of solving crystal structures*. (Cambridge University Press, Cambridge).
85. Emma, P., Akre, R., Arthur, J., Bionta, R., Bostedt, C., Bozek, J., Brachmann, A., Bucksbaum, P., Coffee, R., Decker, F.-J., Ding, Y., Dowell, D., Edstrom, S., A. Fisher, Frisch, J., Gilevich, S., Hastings, J., Hays, G., Hering, Ph., Huang, Z., Iverson, R., Loos, H., Messerschmidt, M., Miahnahri, A., Moeller, S., Nuhn, H.-D., Pile, G., Ratner, D., Rzepiela, J., Schultz, D., Smith, T., Stefan, P., Tompkins, H., Turner, J., Welch, J., White, W., Wu, J., Yocky G., and Galayda1 J. (2010). First lasing and operation of an ångstrom-wavelength free-electron laser. *Nat. Photon.*, **4**, pp. 641.
86. http://www-xfel.spring8.or.jp/SCSSCDR.pdf
87. Altarelli, M., et al. (2006) *XFEL: The European X-ray Free-electron Laser. Technical Design Report DESY 2006-097* (http://xfel.desy.de/tdr/tdr/)
88. Neutze, R., Wouts, R., van der Spoel, D., Weckert, E. and Hajdu, J. (2000). Potential for biomolecular imaging with femtosecond X-ray pulses. *Nature*, **406**, pp. 752.
89. Gaffney, K. J. and Chapman, H. N. (2007). Imaging Atomic Structure and Dynamics with Ultrafast X-ray Scattering. *Science*, **316**, pp. 1444-1448.
90. Singer, A., Vartanyants, I. A., Kuhlmann, M., Düsterer, S., Treusch, R. and Feldhaus, J. (2008). Transverse-Coherence Properties of the Free-Electron-Laser FLASH at DESY. *Phys, Rev. Lett.*, **101**, pp. 254801.
91. Vartanyants, I. A., Singer, A., Mancuso, A. P., Yefanov, O. M., Sakdinawat, A., Liu, Y., Bang, E., Williams, G. J., Cadenazzi, G., Abbey, B., Sinn, H., Attwood, D., Nugent, K. A., Weckert, E., Wang, T., Zhu, D., Wu, B., Graves, C., Scherz, A., Turner, J. J., Schlotter, W.F., Messerschmidt, M., Luning, J., Acremann, Y., Heimann, P., Mancini, D. C., Joshi, V., Krzywinski, J., Soufli, R., Fernandez-Perea, M., Hau-Riege, S., Peele, A. G., Feng, Y., Krupin, O., Möller, S., and Wurth, W. (2011). Coherence Properties of Individual Femtosecond Pulses of an X-Ray Free-Electron Laser, *Phys. Rev. Lett.*, **107**, pp. 144801/1-5.
92. Howells, M. R., Beetz, T., Chapman, H. N., Cui, C., Holton, J. M., Jacobsen, C. J., Kirz, J., Lima, E., Marchesini, S., Miao, H., Sayre, D., Shapiro, D. A., Spence, J. C. H., Starodube D. (2009). An assessment of the resolution limitation due to radiation-damage in x-ray diffraction microscopy. *J. Electron Spectrosc. Relat. Phenom.*, **170,** pp. 4-12.
93. Nave, C. and Garman, E. F. (2005). Towards an understanding of radiation damage in cryocooled macromolecular crystals. *J. Synchrotron Rad.*, **12**, pp. 257.
94. Ackermann, W., et al. (2007). Operation of a Free Electron Laser in the Wavelength Range from the Extreme Ultraviolet to the Water Window. *Nature Photonics*, **1**, pp. 336-342.
95. Wellhoefer, M., Martins, M., Wurth, W., Sorokin, A. A. and Richter, M. (2007). Performance of the monochromator beamline at FLASH. *J. Opt. A: Pure Appl. Opt.*, **9**, pp. 749.

96. Chapman, H.N., et al. (2011). Femtosecond X-ray protein nanocrystallography. *Nature*, **470**, pp. 73-77.
97. Seibert, M., et al. (2011). Single mimivirus particles intercepted and imaged with an X-ray laser. Nature, **470**, pp. 78-81.
98. Bei, M., Borland, M., Cai, Y., Elleaume, P., Gerig, R., Harkay, K., Emery, L., Hutton, A., Hettel, R., Nagaoka, R., Robin, D. and Steier, C. (2010). The Potential of an Ultimate Storage Ring for Future Light Sources. *Nucl. Instrum.Meth. Phys. Res. A*, **622**, pp. 518-535.
99. Mimura, H., Handa, S., Kimura, T., Yumoto, H., Yamakawa, D., Yokoyama, H., Matsuyama, S., Inagaki, K., Yamamura, K., Sano, Y., Tamasaku, K., Nishino, Y., Yabashi, M., Ishikawa, T., and Yamauchi, K. (2010). Breaking the 10 nm barrier in hard-X-ray focusing. *Nature Phys.*, **6**, pp. 122-125.